\documentclass[useAMS,usenatbib]{mnras}
\usepackage{graphicx}    
\usepackage{dcolumn}     
\usepackage{bm}          
\usepackage{amssymb,amsmath}  
\usepackage{color}
\usepackage{hyperref}
\usepackage{cleveref}
\usepackage{natbib}
\usepackage{amsbsy}
\usepackage{overpic}
\usepackage{eso-pic}

\title[Magnification in DES-SV]
      {Weak lensing magnification in the Dark Energy Survey Science Verification data}
\author[M.~Garcia-Fernandez et~al.]{
\parbox{\textwidth}{
\Large
M.~Garcia-Fernandez$^{1}$\thanks{\href{mailto:manuel.garcia-fernandez@ciemat.es}{manuel.garcia-fernandez@ciemat.es}},
E.~Sanchez$^{1}$,
I.~Sevilla-Noarbe$^{1}$,
E.~Suchyta$^{2}$,
E.~M.~Huff$^{3}$,
E.~Gaztanaga$^{4}$,
J.~Aleksi\'c$^{5}$,
R.~Ponce$^{1}$,
F.~J.~Castander$^{4}$,
B.~Hoyle$^{6}$,
T. M. C.~Abbott$^{7}$,
F.~B.~Abdalla$^{8,9}$,
S.~Allam$^{10}$,
J.~Annis$^{10}$,
A.~Benoit-L{\'e}vy$^{8,11,12}$,
G.~M.~Bernstein$^{13}$,
E.~Bertin$^{11,12}$,
D.~Brooks$^{8}$,
E.~Buckley-Geer$^{10}$,
D.~L.~Burke$^{14,15}$,
A. Carnero Rosell$^{16,17}$,
M.~Carrasco~Kind$^{18,19}$,
J.~Carretero$^{4,5}$,
M.~Crocce$^{4}$,
C.~E.~Cunha$^{14}$,
C.~B.~D'Andrea$^{20,21}$,
L.~N.~da Costa$^{16,17}$,
D.~L.~DePoy$^{22}$,
S.~Desai$^{23}$,
H.~T.~Diehl$^{10}$,
T.~F.~Eifler$^{3}$,
A.~E.~Evrard$^{24,25}$,
E.~Fernandez$^{5}$,
B.~Flaugher$^{10}$,
P.~Fosalba$^{4}$,
J.~Frieman$^{10,26}$,
J.~Garc\'ia-Bellido$^{27}$,
D.~W.~Gerdes$^{25}$,
T.~Giannantonio$^{28,29}$,
D.~Gruen$^{14,15}$,
R.~A.~Gruendl$^{18,19}$,
J.~Gschwend$^{16,17}$,
G.~Gutierrez$^{10}$,
D.~J.~James$^{7,30}$,
M.~Jarvis$^{13}$,
D.~Kirk$^{8}$,
E.~Krause$^{14}$,
K.~Kuehn$^{31}$,
N.~Kuropatkin$^{10}$,
O.~Lahav$^{8}$,
M.~Lima$^{16,32}$,
N.~MacCrann$^{33}$,
M.~A.~G.~Maia$^{16,17}$,
M.~March$^{13}$,
J.~L.~Marshall$^{22}$,
P.~Melchior$^{34}$,
R.~Miquel$^{5,35}$,
J.~J.~Mohr$^{36,37,38}$,
A.~A.~Plazas$^{3}$,
A.~K.~Romer$^{39}$,
A.~Roodman$^{14,15}$,
E.~S.~Rykoff$^{14,15}$,
V.~Scarpine$^{10}$,
M.~Schubnell$^{25}$,
R.~C.~Smith$^{7}$,
M.~Soares-Santos$^{10}$,
F.~Sobreira$^{16,40}$,
G.~Tarle$^{25}$,
D.~Thomas$^{20}$,
A.~R.~Walker$^{7}$,
W.~Wester$^{10}$
\begin{center} (The DES Collaboration) \end{center}
}
\vspace{0.4cm}
\\
\parbox{\textwidth}{\rm Author affiliations are listed at the end of this paper.}
}

\date{Accepted XXX. Received YYY; in original form ZZZ}
\pubyear{2017}
\begin{document}

\label{firstpage}
\pagerange{\pageref{firstpage}--\pageref{lastpage}}
\maketitle

\begin{abstract}
In this paper the effect of weak lensing magnification on galaxy number counts is studied by cross-correlating the positions of two galaxy samples, separated by redshift, using the Dark Energy Survey Science Verification dataset. This analysis is carried out for galaxies that are selected only by its photometric redshift. An extensive analysis of the systematic effects, using new methods based on simulations is performed, including a Monte Carlo sampling of the selection function of the survey.
\end{abstract}

\begin{keywords}
methods: data analysis -- techniques: photometric -- gravitational lensing: weak -- large-scale structure of the Universe 
\end{keywords}
\section{Introduction}
\label{sec:intro}

Weak gravitational lensing of distant objects by the nearby large-scale structure of the Universe is a powerful probe of cosmology \citep{2001PhR...340..291B,2006glsw.conf.....M,2008ARNPS..58...99H,2041-8205-723-1-L13,Weinberg201387,2015RPPh...78h6901K} with two main signatures: magnification and shear.

Magnification is due to the gravitational bending of the light emitted by distant sources by the matter located between those sources and the observer \citep{1989Sci...245..824B}. This leads to an isotropic observed size enlargement of the object while the surface brightness is conserved \citep{1992ARA&A..30..311B}, modifying three observed properties of the sources: size, magnitude and spatial density. The change of spatial density of galaxies due to gravitational lensing is known as number count magnification and arises from the increase of the observed flux of the background galaxies, allowing the detection of objects that, in the absence of lensing, would be beyond the detection threshold \citep{1992S&W....31..459B}. Magnification is dependent on the mass of the dark matter content along the line of sight to the source \citep{1992RvMA....5..259B,1992LNP...406..345B,1995A&A...303..643B,1995AIPC..336..307B}. Therefore, its effect is not homogeneous and is spatially correlated with the location of lens galaxies and clusters, which are biased tracers of the dark matter field \citep{1978MNRAS.183..341W,1984ApJ...284L...9K}.

Since magnification and shear are complementary effects of the same physical phenomenon, they depend on the same cosmological parameters, but in a slightly different manner. Thus, some degeneracies are broken on parameter constraints (e.g. at the $\Omega_M-\sigma_8$ plane) when combining magnification with shear-shear correlations \citep{2010MNRAS.401.2093V}. Nevertheless, the major power of the combination of both methods is that they are sensitive to different sources of systematic errors. For example, number count magnification is independent of those systematic effects caused by shape determination, although it suffers from selection effects \citep{2015MNRAS.454.3121M}. This constitutes a powerful feature that can be exploited to minimize systematic effects on a possible combination of magnification with galaxy-shear (gg-lensing) since both measurements are produced by the convergence field.

Extensive wide-field programs have allowed accurate measurements of weak lensing effects. Previous magnification measurements involve the use of very massive objects as lenses, such as luminous red galaxies (LRGs) and clusters \citep{1995AIPC..336..320B,2014MNRAS.440.3701B,2014MNRAS.439.3755F,2016MNRAS.457.3050C}, or high redshift objects as sources, such as Lyman break galaxies (LBGs; \cite{2009A&A...507..683H,2012MNRAS.426.2489M}) quasars \citep{1979ApJ...227...30S,1989Natur.339..106H,1990A&A...240...11F,1993A&A...268....1B,2002A&A...386..784M,0004-637X-589-1-82,0004-637X-633-2-589} and sub-mm sources \citep{2011MNRAS.414..596W} to improve signal-to-noise ratio. In addition to the number count technique used in this paper, other observational effects produced by magnification have been measured as well: the shift in magnitude \citep{2010MNRAS.405.1025M}, flux \citep{2011MNRAS.411.2113J} and size \citep{2041-8205-780-2-L16}.

Lyman break galaxies and quasars have demonstrated to be a very effective population of background samples to do  magnification  studies  due  to  its  high  lensing  efficiency. However deep surveys or large areas are needed to reach a significant number of these objects.  Thus,  shallow  or  small  area  surveys  require  the selection of a more numerous population of source galaxies to allow the measurement of the magnification signal.

In this paper, the magnification signal is measured using the number-count technique on the Dark Energy Survey\footnote{www.darkenergysurvey.org} (DES) Science Verification data. All observed galaxies, selected only with photometric redshifts, are used both as lenses and sources. This procedure simplifies the analysis as no addition processing or selections are needed to construct the sample, as in the dropout technique used for LBG selection. This alternative way to select galaxies is different to what is found on previous works, and provides a more numerous source sample. This allows the detection of magnification on small area surveys --such as the DES Science Verification data--, but the main power of this methodology resides on photometric surveys with large areas such as LSST\footnote{www.lsst.org} and the final footprint of DES, with 5000 deg$^2$. The increase on the density of sources on large areas provides a huge number of total sources, reducing dramatically the shot-noise.
\newline

In addition, this paper presents an extensive test for systematic effects. New techniques, based on simulations specially developed for this purpose are used, including a Monte Carlo sampling method to model the selection function of the survey.

The structure of the paper is as follows: in \autoref{sec:mag_theory} the theory behind magnification is summarized. The steps leading to a detection are described in \autoref{sec:method}, and \autoref{sec:data_sample} describes the data sample. The methodology is validated in \autoref{sec:simulation} with a study on N-body simulations. The analysis of the data sample is made in \autoref{sec:data_analysis}, concluding in \autoref{sec:conclusions}.

\section{Number Count Magnification}
\label{sec:mag_theory}
Number count magnification can be detected and quantified by the deviation of the expected object counts in the positional correlation of a foreground and a background galaxy sample \citep{1979ApJ...227...30S}. These galaxy samples, in absence of magnification, are uncorrelated if their redshift distributions have a negligible overlap. In this section, the formalism that will quantify its effect on this observable is presented.

The observed two-point angular cross-correlation function between the {\it i-} and {\it j-}th redshift bins, including magnification, is defined as \citep{1995A&A...298..661B}
\begin{equation}
\omega_{ij}(\theta) = \langle \delta_O(\boldsymbol{\hat n},z_i,f_i)\delta_O(\boldsymbol{\hat n'},z_j,f_j)\rangle_\theta,
\end{equation}
where $\theta$ is the angle subtended by the two direction vectors $\boldsymbol{\hat n},\boldsymbol{\hat n'}$ and the observed density contrast ($\delta_O$) is
\begin{equation}
\delta_O(\boldsymbol{\hat n},z_i,f_i) = \delta_g(\boldsymbol{\hat n},z_i)+\delta_\mu(\boldsymbol{\hat n},z_i,f_i);
\label{eq:delta0}
\end{equation}
where $\delta_g$ describes the fluctuations due to the intrinsic matter clustering at redshift $z_i$ and $\delta_\mu$ incorporates the fluctuations from magnification effects at a flux cut $f_i$.

The galaxy density contrast in the linear bias approximation is (\cite{1994MNRAS.267.1020P}; \cite{2015MNRAS.448.1389C})
\begin{equation}
\delta_g(\boldsymbol{\hat n},z_i) = b_i\delta_M(\boldsymbol{\hat n},z_i)
\label{eq:deltaM}
\end{equation}
with $b_i$ the galaxy-bias at redshift $z_i$ and $\delta_M$ the intrinsic matter density contrast.

Following the approach used by \cite{2001PhR...340..291B} and \cite{2003A&A...403..817M}, the magnification density contrast on the sky in direction $\boldsymbol{\hat n}$  is defined as
\begin{equation}
\delta_\mu(\boldsymbol{\hat n},z,f_\mu) = \frac{N_\mu(\boldsymbol{\hat n},z,f_\mu)}{N_0(\boldsymbol{\hat n},z,f_0)}-1.
\end{equation}
Here $N_0(\boldsymbol{\hat n},z,f_0)$ is the unlensed cumulative number count of sources located at redshift $z$, that is, the number of sources with observed flux greater than the threshold $f_0$, while, $N_\mu(\boldsymbol{\hat n},z,f_\mu)$ is the lensed cumulative number count, affected by magnification.

Magnification by gravitational lenses increases the observed flux of background galaxies allowing one to see fainter sources changing the effective flux cut from $f_0$ to $f_\mu = f_0/\mu$. At the same time it stretches the solid angle behind the lenses, reducing the surface density of sources down to $N_\mu=N_0/\mu$ \citep{1989ApJ...339L..53N}. Thus the density contrast may be rewritten as
\begin{equation}
\delta_\mu(\boldsymbol{\hat n},z,f_\mu) = \frac{N_\mu(\boldsymbol{\hat n},z,f_\mu)}{\mu N_\mu(\boldsymbol{\hat n},z,\mu f_\mu)}-1.
\label{eq:denscont}
\end{equation}
The cumulative number count can be locally parametrized as
\begin{equation}
N_\mu(\boldsymbol{\hat n},z,f_\mu)=A\left(\frac{f_\mu}{f_*}\right)^{\alpha(f_{\mu})}
\end{equation}
where $A,f_*$ are constant parameters and $\alpha(f_\mu)$ is a function of the flux limit. Substituting this parametrization into \autoref{eq:denscont}: 
\begin{equation}
\delta_\mu(\boldsymbol{\hat n},z,f_\mu) = \mu^{-\alpha(f_\mu)-1}-1.
\end{equation}
Taking the weak lensing approximation, $\mu \simeq 1+2\kappa$ with $\kappa\ll1$, where $\kappa$ corresponds to the lensing convergence of the field \citep{1992A&A...259..413B}, and converting from fluxes to magnitudes, the previous equation becomes \citep{1993LIACo..31..217N}
\begin{equation}
\delta_\mu(\boldsymbol{\hat n},z,m) = 2\kappa(\boldsymbol{\hat n},z)\left[\alpha(m)-1\right]
\label{eq:deltamu}
\end{equation}
with
\begin{equation}
\alpha(m) = 2.5\frac{d}{dm}[\log N_{\mu}(m)].
\label{eq:alpha}
\end{equation}
The convergence $\kappa$ is defined as \citep{1992ARA&A..30..311B,2001PhR...340..291B}
\begin{equation}
\kappa(\boldsymbol{\hat n},z) = \int\limits_0^z dz'\frac{r(z')[r(z)-r(z')]}{r(z)}\nabla_\perp^2\Phi[r(z'),\boldsymbol{\hat n}],
\end{equation}
where $r(z)$ is the radial comoving distance at redshift $z$, $\nabla^2_\perp$ is the Laplacian on the coordinates of the plane transverse to the line of sight and $\Phi$ is the gravitational potential. Assuming that the gravitational potential and the matter density may be written as the sum of an homogeneous term plus a perturbation ($\Phi=\bar\Phi+\delta_\Phi$ and $\rho=\bar\rho+\delta_M$ respectively) the Poisson equation can be written as:
\begin{equation}
\nabla^2\Phi(r,\boldsymbol{\hat n}) = \nabla^2\delta_\Phi(r,\boldsymbol{\hat n}) = 4\pi Ga^2\bar\rho\delta_M(r,\boldsymbol{\hat n}),
\end{equation}
where $a=1/(1+z)$ is the scale factor. Expressing the matter density as a function of the critical matter density at present, this leads to \citep{1989ApJ...344..637G}
\begin{equation}
\nabla^2_\perp\Phi(r,\boldsymbol{\hat n}) = \frac{3H_0^2}{2ac^2}\Omega_M^0\delta_M(r,\boldsymbol{\hat n}),
\end{equation}
with $\delta_g$ the galaxy density contrast, $H_0$ the Hubble constant and $c$ the speed of light.

Combining Equations \ref{eq:delta0}, \ref{eq:deltaM} and \ref{eq:deltamu} it is straightforward to arrive at \citep{PhysRevD.76.103502,PhysRevD.77.023512,PhysRevD.77.063526}:

\begin{subequations}
\begin{eqnarray}
\omega_{ij}(\theta) =& &\langle b_ib_j\delta_M(\boldsymbol{\hat n},z_i)\delta_M(\boldsymbol{\hat n'},z_j)\rangle_\theta\label{eq:4a}\\ 
+&&\langle b_i\delta_M(\boldsymbol{\hat n},z_i)\delta_\mu(\boldsymbol{\hat n'},z_j,m_j)\rangle_\theta\label{eq:4b}\\ 
+&&\langle b_j\delta_M(\boldsymbol{\hat n'},z_j)\delta_\mu(\boldsymbol{\hat n},z_i,m_i)\rangle_\theta\label{eq:4c}\\ 
+&&\langle\delta_\mu(\boldsymbol{\hat n},z_i,m_i)\delta_\mu(\boldsymbol{\hat n'},z_j,m_j)\rangle_\theta.\label{eq:4d}
\end{eqnarray}
\end{subequations}
If it is assumed that $z_i<z_j$ where $z_i$ are the lens redshift bins and $z_j$ the source redshift bins, the only terms that are non-vanishing, assuming well determined redshifts, are Equations \ref{eq:4b} and \ref{eq:4d}, where the last term is subleading, resulting  \citep{2003A&A...403..817M}:
\begin{eqnarray}
\omega_{ij}(\theta)= &b_i[\alpha(m_j)-1]\frac{3H_0^2\Omega_M^0}{c^2}\nonumber\\
\times&\int\limits_0^\infty dz_i'\frac{\phi_i(z_i')}{1+z_i'}\int\limits_{z_i'}^\infty dz_j'\phi_j(z_j')\frac{r(z_i')[r(z_j')-r(z_i')]}{r(z_j')}\\
\times&\int\limits_0^\infty\frac{dkk}{2\pi}P_{M}(k,z_i')J_0(k\theta r(z_i')), \nonumber
\label{eq:wijfinal}
\end{eqnarray}
where $P_M$ is the matter power spectrum, $J_0$ is zero-th order Bessel function and $\phi_i,\phi_j$ are the redshift distribution of the lens and source sample respectively. A short-hand way to express the two point angular cross-correlation function due to magnification between a lens sample (L) and a source sample (S) with magnitude cut $m_j$  is
\begin{equation}
\omega_{LS_j}(\theta) = b_L[\alpha_S(m_j)-1]\omega_0(\theta).
\label{eq:shorthand}
\end{equation}
Here $b_L$ is the galaxy-bias of the lenses, $\alpha_S(m_j)$ the number count slope of the sources given by \autoref{eq:alpha} and $\omega_0(\theta)$ is the angular correlation function of the projected mass on the lens plane, that depends only on the cosmological parameters. The number count slope is evaluated at the threshold magnitude $m_j$, that is, the upper magnitude cut imposed on the $j$-th source sample.

\section{Measuring Magnification through Number Count}
\label{sec:method}
By inspection of Equations \ref{eq:alpha} and \ref{eq:shorthand} and the gravitational lens equation \citep{1992ARA&A..30..311B}, three key properties can be deduced that are intrinsic to magnification:

\begin{itemize}
  \item A non-zero two-point angular cross-correlation $\omega_{LS_j}$ appears between two galaxy samples at redshifts $z_{S_j} > z_L$ for those cases in which the slope $\alpha_S(m_j) \neq 1$ (magnification signal hereafter).
  \item The amplitude of the magnification signal evolves with the slope of the faint end of the number count distribution of the source sample and, assuming a \cite{1976ApJ...203..297S} luminosity function, eventually it reaches zero and becomes negative.
  \item For a given value of the number count slope, the signal strength is independent of the photometric band used (i.e. it is achromatic).
\end{itemize}
The steps towards a measurement of magnification via the number count technique in a photometric survey can be summarized as follows:
\begin{enumerate}
	\item Split the data sample into two well-separated photo-z bins, termed lens and source.
	Splitting must be done minimizing the overlap between the true redshift distributions of the samples. Otherwise, by \autoref{eq:4a}, an additive signal is introduced.
	\item For each photometric band, define several subsamples from the source sample using different values for the maximum (threshold) magnitude. This is made in order to trace the evolution of the amplitude of the magnification signal with the number count slope (see \autoref{eq:alpha}).
	\item Compute the two-point angular cross-correlation function between the unique common lens sample and each source subsample for each band.
\end{enumerate}
Once the two-point angular correlation function have been measured, They can be compared with theoretical predictions as described in \autoref{sec:mag_theory}. allowing the desired determination.

\section{The Data Sample}
\label{sec:data_sample}
The Dark Energy Survey (DES; \cite{2005IJMPA..20.3121F}) is a photometric galaxy survey that uses the Dark Energy Camera (DECam; \cite{DIEHL20121332,2015AJ....150..150F}), mounted at the Blanco Telescope, at the Cerro Tololo Interamerican Observatory. The survey will cover about $5000 \mbox{ deg}^2$ of the southern hemisphere, imaging around $3\times10^8$ galaxies in 5 broad-band filters ({\it grizY}) at limiting magnitudes $g<24.6$, $r<24.1$, $i<24.3$, $z<23.9$. The sample used in this analysis corresponds to the Science Verification (DES-SV) data, which contains several disconnected fields. From the DES SVA1-Gold\footnote{des.ncsa.illinois.edu/releases/SVA1} main galaxy catalog \citep{2016MNRAS.455.4301C}, the largest contiguous field is selected, the SPT-E. Regions with declination $ < -61^{\circ}$ are removed in order to avoid the Large Magellanic Cloud. {\scshape Modest\_class} is employed as star-galaxy classifier \citep{0004-637X-801-2-73}.

The following colour cuts are made in order to remove outliers in colour space:
\begin{itemize}
	\item $-1 < g-r < 3$,
	\item $-1 < r-i < 2$,
	\item $-1 < i-z < 2$;
\end{itemize}
where {\it g}, {\it r}, {\it i}, {\it z} stand for the corresponding {\scshape mag\_auto} magnitude measured by {\scshape SExtractor} \citep{1996A&AS..117..393B}.

Regions of the sky that are tagged as bad, amounting to four per cent of the total area, are removed. An area of radius 2 arcminutes around each 2MASS star is masked to avoid stellar halos (\cite{2005MNRAS.361.1287M}; \cite{0004-637X-633-2-589}).

The DES Data Management \citep{2011arXiv1109.6741S,2012ApJ...757...83D,2012SPIE.8451E..0DM} produces a {\scshape mangle}\footnote{http://space.mit.edu/$\sim$molly/mangle/} \citep{2008MNRAS.387.1391S}  magnitude limit mask that is later translated to a $N_{\rm side}=4096$ {\scshape HEALPix}\footnote{healpix.jpl.nasa.gov} \citep{2005ApJ...622..759G} mask. Since the {\scshape HEALPix} mask is a division of the celestial sphere on romboid-like shaped pixels with the same area, to avoid boundary effects due to the possible mismatch between the {\scshape mangle} and {\scshape HEALPix} masks, each pixel is required to be totally inside the observed footprint as determined by {\scshape mangle}, by demanding
\begin{itemize}
	\item $r_{\rm fracdet}=1$,
	\item $i_{\rm fracdet}=1$,
	\item $z_{\rm fracdet}=1$;
\end{itemize}
where $r_{\rm fracdet},i_{\rm fracdet},z_{\rm fracdet}$ is the fraction of the pixel lying inside the footprint for {\it r}, {\it i}, {\it z} bands respectively.

\begin{figure}
\begin{flushright}
\begin{overpic}[width=0.45\textwidth]{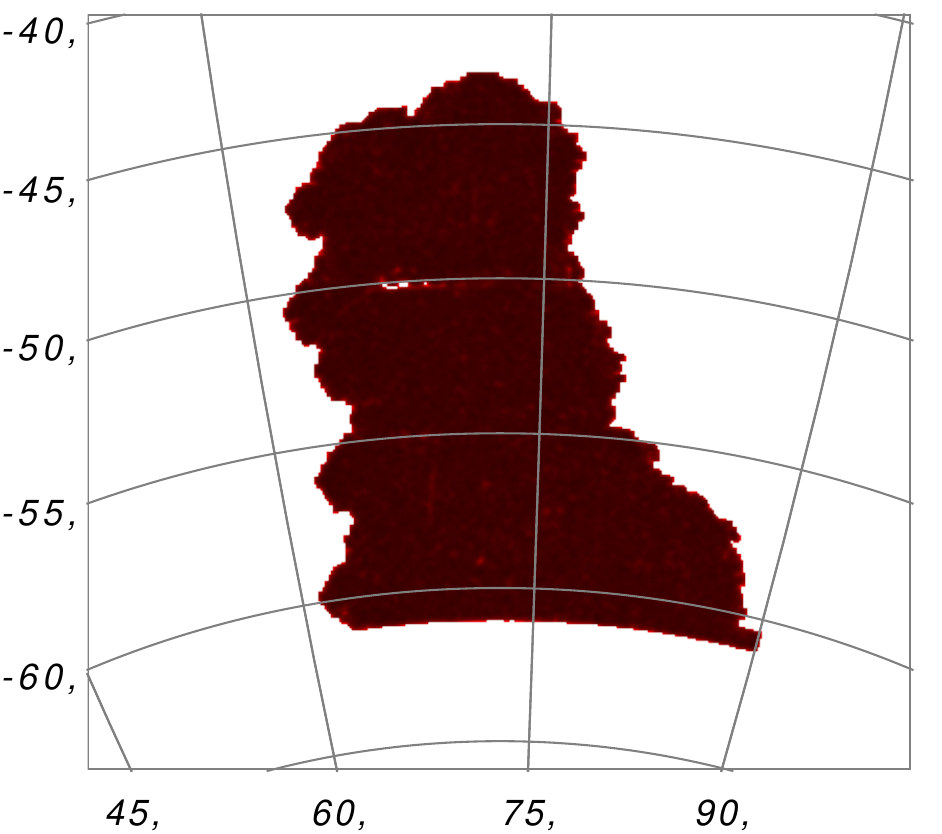}
\put(85,0){RA [$^\circ$]}
\put(-5,70){\rotatebox{90}{DEC [$^\circ$]}}
\end{overpic}
\end{flushright}
\caption{Final footprint of the DES SPT-E region after all masking is applied.}
\label{fig:footprint}
\end{figure}

Depth cuts are also imposed on the {\it riz}-bands in order to have uniform depth when combined with the magnitude cuts. These depth cuts are reached by including only the regions that meet the following conditions:
\begin{itemize}
	\item $r_{\rm lim} > 23.0$,
	\item $i_{\rm lim} > 22.5$,
	\item $z_{\rm lim} > 22.0$;
\end{itemize}
where $r_{\rm lim}, i_{\rm lim},z_{\rm lim}$ stand for the magnitude limit in the corresponding band, that is, the faintest magnitude at which the flux of a galaxy is detected at 10$\sigma$ significance level. The resulting footprint, as shown in \autoref{fig:footprint}, after all the masking cuts amounts to $121 \mbox{ deg}^2$.

\begin{figure}
\includegraphics[width=0.5\textwidth]{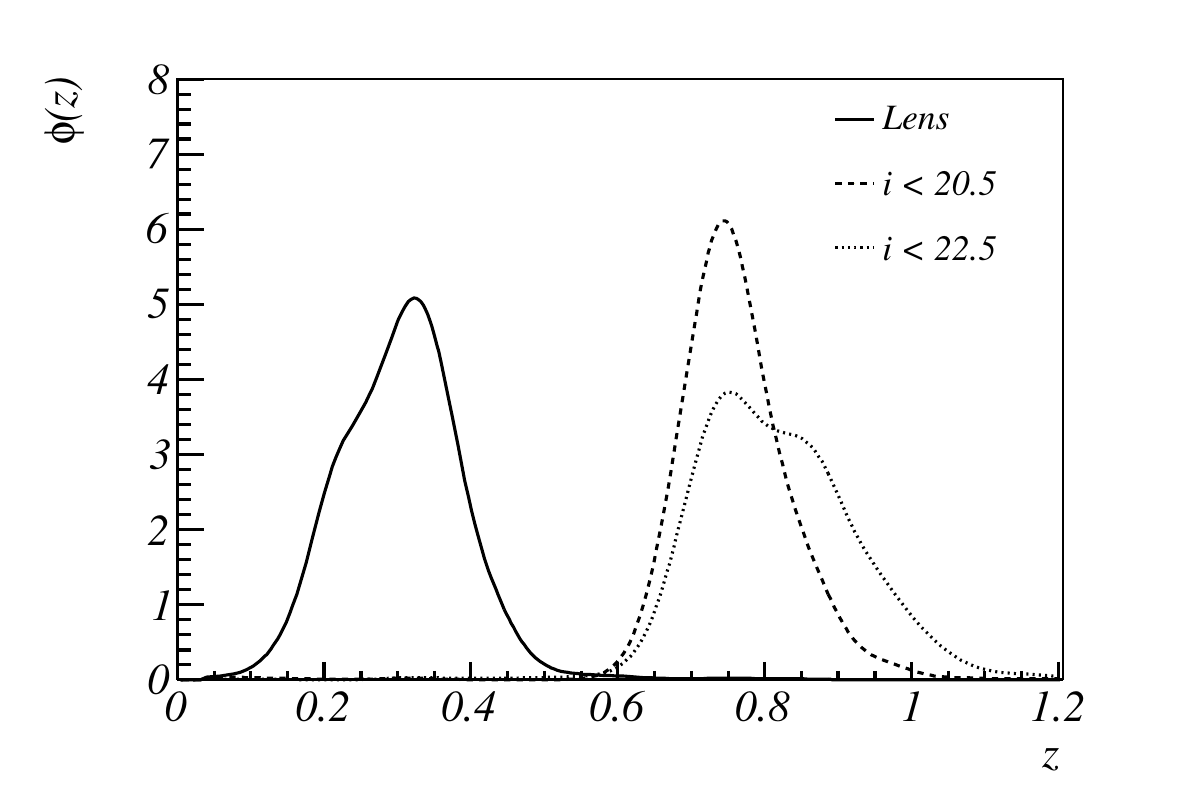}
\caption{Redshift distributions from the stacking of the TPZ probability distribution functions for the lens and two {\it i}-band sub-samples of the source.}
\label{fig:stacking}
\end{figure}

Photometric redshifts (photo-z) have been estimated using different techniques. In particular, the fiducial code used in this work employs a machine-learning algorithm (random forests) as implemented by TPZ \citep{2013MNRAS.432.1483C}, which was shown to perform well on SV data \citep{2014MNRAS.445.1482S}. The redshifts of the galaxies are defined according to the mean of the probability density functions given by TPZ ($z_{\rm ph}$). Other methods are also employed to demonstrate that the measured two-point angular cross-correlation are not a feature induced by TPZ (see \autoref{sec:sys}).

\subsection{Lens sample}
A unique lens sample is defined by the additional photo-z and magnitude cuts:
\begin{itemize}
	\item $0.2 < z_{\rm ph} < 0.4$;
	\item $18.0 < i < 22.5$.
\end{itemize}
These requirements are imposed in order to be compatible with the first redshift bin of the so called `benchmark sample' \citep{2016MNRAS.455.4301C}. Note that the {\scshape mag\_auto} cut along with the previous $i$-band depth cut guarantees uniformity \citep{2016MNRAS.455.4301C}.

\subsection{Source sample}
Three source samples are defined, one per band:
\begin{itemize}
	\item R: $0.7 < z_{\rm ph} < 1.0$ and $r<23.0$;
	\item I: $0.7 < z_{\rm ph} < 1.0$ and $i<22.5$;
	\item Z: $0.7 < z_{\rm ph} < 1.0$ and $z<22.0$.
\end{itemize}
Following the same approach we used on the lens, defined over the `benchmark' sample, the {\scshape mag\_auto} cut along with the previously defined depth cuts also guarantee uniformity on the corresponding band.

Within each R, I, Z source sample five sub-samples that map the magnitude evolution are defined,
\begin{itemize}
	\item $\rm R_1$: $r<21.0$; $\rm R_2$: $r<21.5$; $\rm R_3$: $r<22.0$; $\rm R_4$: $r<22.5$; $\rm R_5$: $r<23.0$.
	\item $\rm I_1$: $i<20.5$; $\rm I_2$: $i<21.0$; $\rm I_3$: $i<21.5$; $\rm I_4$: $i<22.0$; $\rm I_5$: $i<22.5$.
	\item $\rm Z_1$: $z<20.0$; $\rm Z_2$: $z<20.5$; $\rm Z_3$: $z<21.0$; $\rm Z_4$: $z<21.5$; $\rm Z_5$: $z<22.0$.
\end{itemize}
Here $\rm S_j$ with $\rm j=1,2,3,4,5$ are the sub-samples of sample S with $\rm S\in \{R,I,Z\}$. In \autoref{fig:stacking}, the redshift distributions of the lens and source sample are shown. Note that the sub-samples $\rm R_5, I_5, Z_5$ are equal to $\rm R, I , Z$ respectively.

The {\it g}-band is not used on this analysis because when the same approach is followed and a uniform sample is defined in that band, the number of galaxies of the lens and source samples decrease dramatically. This increases the shot noise preventing the measurement of number count magnification.
\section{Application to a simulated galaxy survey}
\label{sec:simulation}
\begin{figure}
\includegraphics[width=0.5\textwidth]{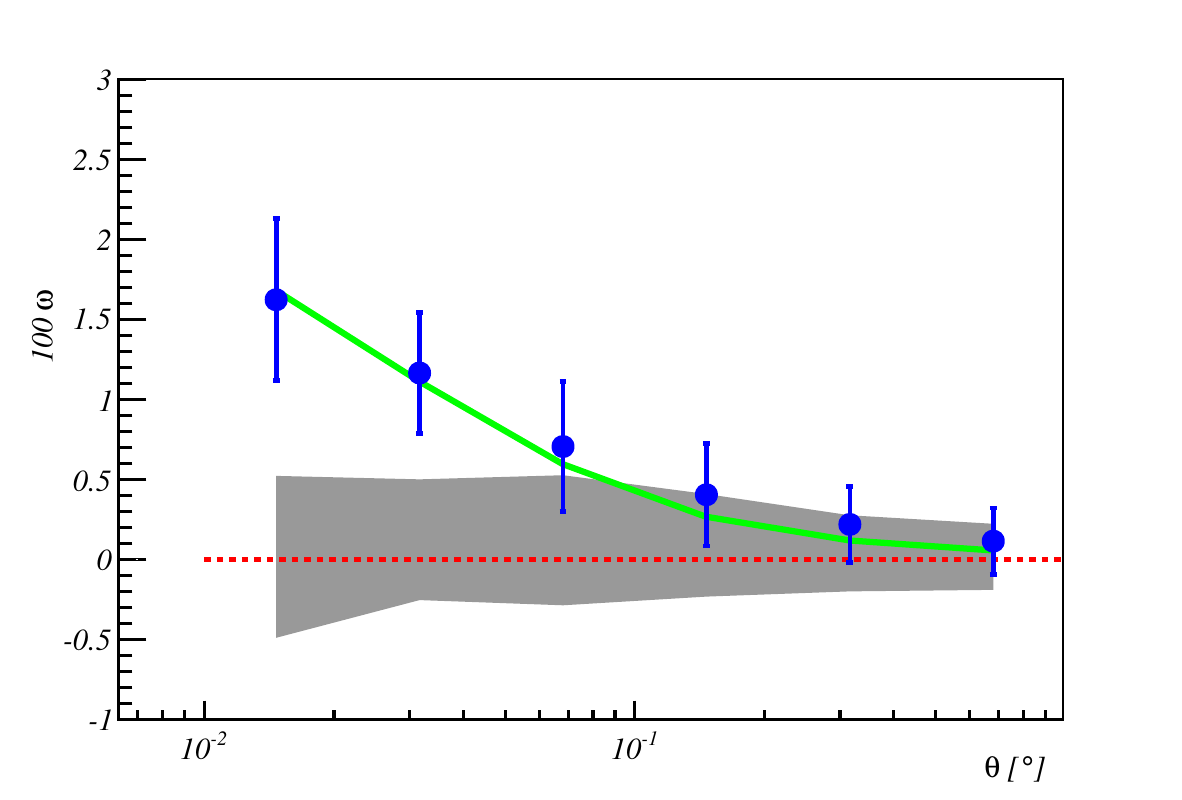}
\caption{Two-point angular cross-correlation function for the MICE simulation (sample $i < 21.5$). With magnification (dots) and without (grey shade), versus that expected from the MICE cosmological parameters, both with magnification (solid line) and without (dashed line), the latter being zero.}
\label{fig:MICE}
\end{figure}
In order to test the methodology described above in a controlled environment, isolated from any source of systematic error, it is applied to a simulated galaxy sample, in particular {\scshape MICECAT v1.0}.
This mock is the first catalog release of the N-body simulation MICE-GC\footnote{www.ice.cat/mice} \citep{2015MNRAS.447.1319F,2015.2987F,2015MNRAS.453.1513C}. It assumes a flat $\Lambda$CDM Universe with cosmological parameters $\Omega_M=0.25, \Omega_b=0.044, h=0.7$ and $\sigma_8 = 0.8$, using a light-cone that spans one eighth of the celestial sphere. Another advantage of using these simulations is the possibility of studying specific systematic effects, as described in \autoref{sec:sys}.

Among other properties, MICE-GC provides lensed and unlensed coordinates, true redshift (including redshift space distortions) and DES-$griz$ unlensed magnitudes for the simulated galaxies, along with convergence and shear.
Conversion from unlensed magnitudes to lensed magnitudes can be done by applying $m_\mu = m_0-2.5\log_{10}(1+2\kappa)$.

Having two sets of coordinates and magnitudes, one in a `universe' with magnification and another without magnification, allows us to follow the methodology described in \autoref{sec:method} for both cases,
serving as a test-bench to measure the sensitivity of the method to the magnification effect. In order to have a fiducial function with as little statistical uncertainty as possible, the full 5000 deg$^2$ of the MICE simulation are used. To match as much as possible the conditions of the DES-SV data, the magnitude cuts described in \autoref{sec:data_sample} are applied to the lens and source samples. The covariance matrices of data (see \autoref{sec:data}) are used, in order to match the errors in the DES-SV sample.

In \autoref{fig:MICE}, the results of the magnification analysis in the MICE simulation for the cases with and without magnification can be seen compared with the theoretical expectations. The methodology used in this work clearly allows us to distinguish both cases for a data-set similar to that of the DES-SV data. Nevertheless, results obtained with the MICE simulation can not be directly extrapolated to SV data to estimate the expected significance because the density of galaxies on the simulation is a factor $\sim3$ smaller than on the SV data. Also, the luminosity function of the simulation is slightly different from the DES data, which has a direct impact on the number count slope and, consequently, on the amplitude of the measured signal.
\section{Data Analysis}
\label{sec:data_analysis}

The analysis of the SV data is described here, showing first the detection of the magnification signal followed by the study and correction of systematic effects.

\subsection{Signal detection}
\label{sec:data}
To estimate the cross-correlation functions, the tree-code {\scshape TreeCorr}\footnote{github.com/rmjarvis/TreeCorr} \citep{2004MNRAS.352..338J} and the Landy-Szalay estimator \citep{1993ApJ...412...64L} are used demanding six logarithmic angular bins:
\begin{equation}
\omega_{\rm LS_j}(\theta) = \frac{D_{\rm L}D_{\rm S_j}(\theta)-D_{\rm L}R_{\rm S_j}(\theta)-D_{\rm S_j}R_{\rm L}(\theta)}{R_{\rm L}R_{\rm S_j}(\theta)}+1,
\label{eq:landyszalay}
\end{equation}
where $D_{\rm L}D_{\rm S_j}(\theta)$ is the number of pairs from the lens data sample L and the source data sub-sample $\rm S_j$ separated by an angular distance $\theta$ and $D_{\rm L}R_{\rm S_j}(\theta)$, $D_{\rm S_j}R_{\rm L}(\theta)$, $R_{\rm L}R_{\rm S_j}(\theta)$ are the corresponding values for the lens-random, source-random and random-random combinations normalized by the total number of objects on each sample.

Catalogs produced with {\scshape Balrog}\footnote{github.com/emhuff/Balrog} \citep{2016MNRAS.457..786S} are used as random samples. The {\scshape Balrog} catalogs are DES-like catalogs, where
no intrinsic magnification signal has been included. The {\scshape Balrog} software generates images of fake objects, all with zero convergence $\kappa$, that are embedded into the DES-SV coadd images
(convolving the objects with the measured point spread function, and applying the measured photometric calibration). Then {\scshape SExtractor} was run on them, using the same DES Data Management configuration parameters used for the image processing. The positions for the simulated objects were generated randomly over the celestial sphere, meaning that these positions are intrinsically unclustered. Hence, the detected {\scshape Balrog} objects amount to a set of random points, which sample the survey detection probability. For a full description and an application to the same measurement as in \cite{2016MNRAS.455.4301C} see \cite{2016MNRAS.457..786S}. This is the first time that this extensive simulation is used to correct for systematics.

The same cuts and masking of the data sample (\autoref{sec:data_sample}) are also applied to the the {\scshape Balrog} sample. A re-weighting following a nearest-neighbours approach was applied to {\scshape Balrog} objects in order to follow the same magnitude distribution of the DES-SV data on both lens and sources.

\begin{figure}
\includegraphics[width=0.5\textwidth,trim={0 0 0 2cm},clip]{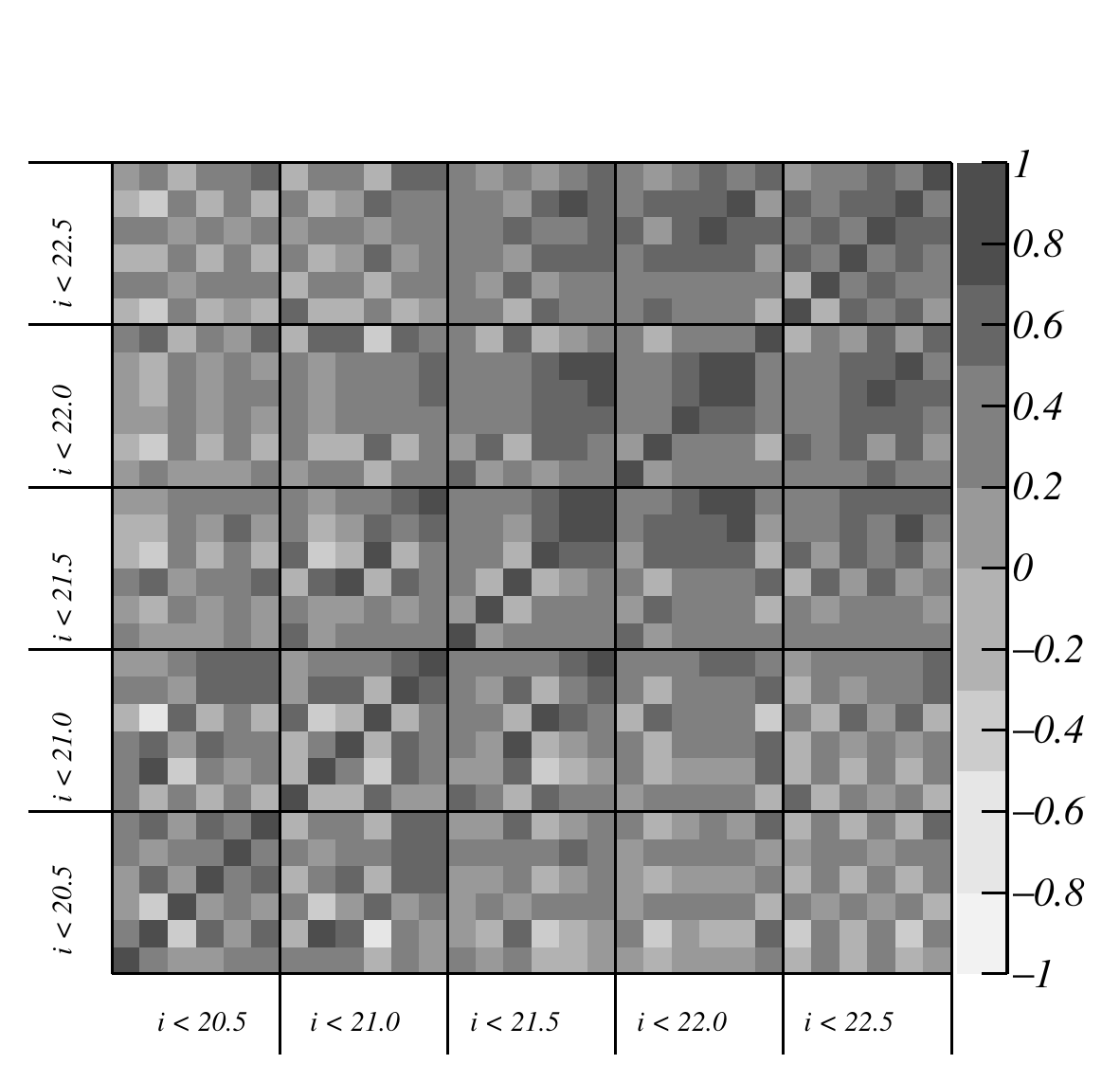}
\caption{Covariance matrix of the {\it i}-band rescaled by the value of the diagonal ($C_{ij}/\sqrt{C_{ii}C_{jj}}$). Each box is the part of the matrix corresponding to the samples labeled at the axis whereas the bins within each box stand for the angular values of the correlation function.}
\label{fig:cov_matrix}
\end{figure}

A covariance matrix is computed for each band by jack-knife re-sampling the data taking into account the correlations between the different magnitude cut within each band
\begin{eqnarray}
C_{\rm S}(\omega_{\rm LS_i}(\theta_\eta);\omega_{\rm LS_j}(\theta_\nu)) = \frac{N_{\rm JK}}{N_{\rm JK}-1}\\
\times \sum\limits_k^{N_{\rm JK}}[\omega_{\rm LS_i}^k(\theta_\eta)-\omega_{\rm LS_i}(\theta_\eta)][\omega_{\rm LS_j}^k(\theta_\nu)-\omega_{\rm LS_j}(\theta_\nu)]\nonumber,
\end{eqnarray}
where $\omega^k_{\rm LS_j}$ stands for the cross-correlation of the $k$-th jack-knife re-sample and $\omega_{\rm LS_j}$ is the cross-correlation of the full sample. The $N_{\rm JK}= 120$ jack-knife regions are defined by a $k$-means algorithm \citep{macqueen1967some} using Python's machine learning library {\scshape scikit-learn}\footnote{scikit-learn.org} \citep{scikit-learn}. In order to get $N_{\rm JK}$ regions with equal area, the algorithm is trained on a uniform random sample following the footprint of the data demanding $N_{\rm JK}$ centres. The regions used on the re-sampling are composed by the Voronoi tessellation defined by these centres. These matrices trace the angular covariance as well as the covariances between functions within each band. No covariance between bands is considered, since each band is treated independently on this work. The reduced covariance matrix of the {\itshape i}-band is displayed at \autoref{fig:cov_matrix}. The behavior is similar for the other bands.

\begin{figure*}
\includegraphics[width=\textwidth,trim={0 2.3cm 0 3.5cm},clip]{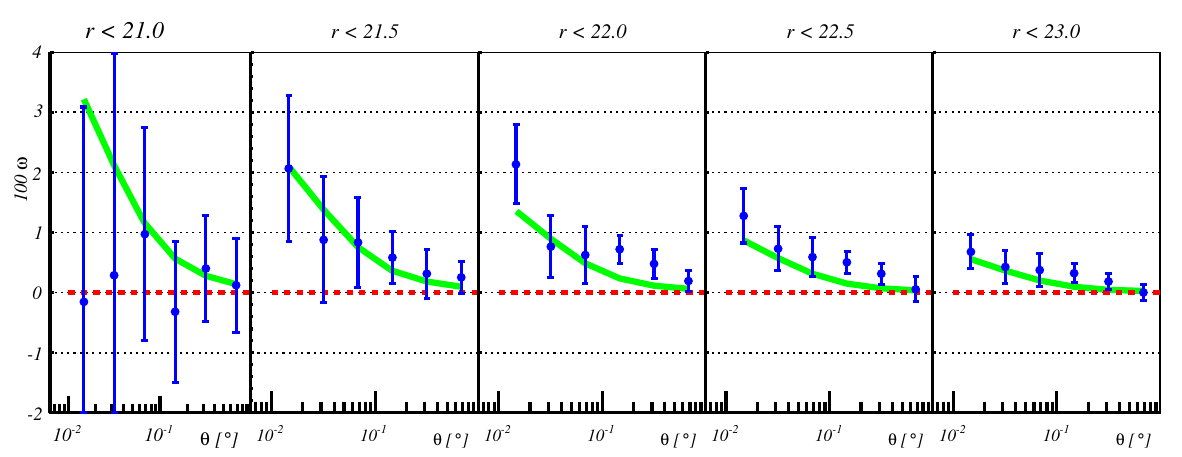}
\includegraphics[width=\textwidth,trim={0 2.3cm 0 3.5cm},clip]{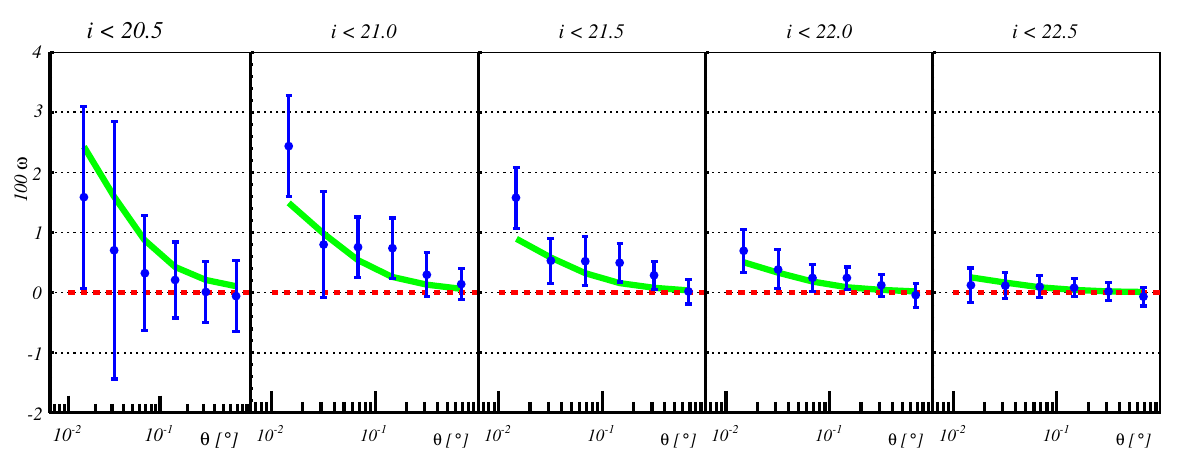}
\includegraphics[width=\textwidth,trim={0 2.3cm 0 3.5cm},clip]{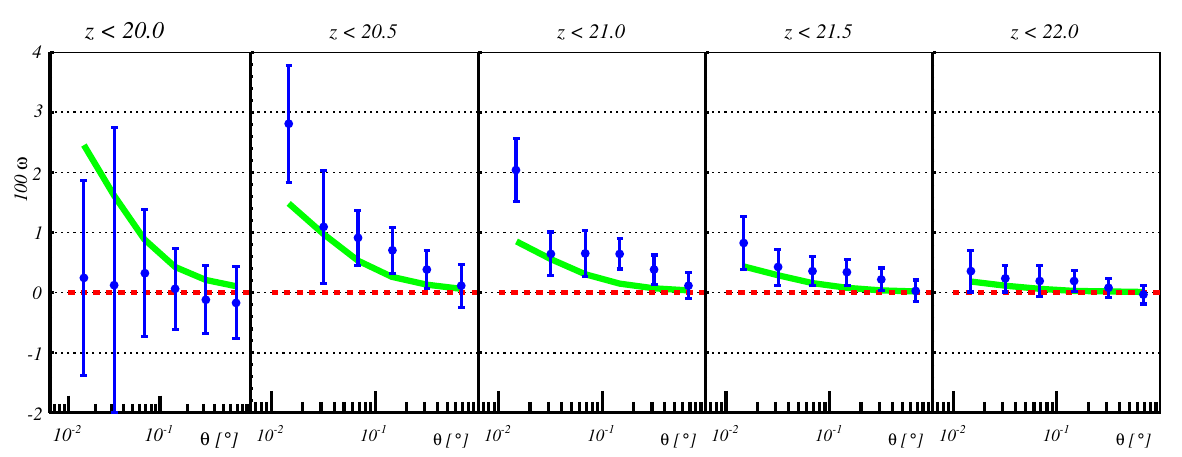}
\caption{Measured cross-correlation functions (dots) of the lens sample with each source sample for the DES SVA1-Gold data using {\scshape Balrog} randoms. Each row corresponds to one of the R, I, Z source samples. Within each row, each sub-panel shows the cross-correlation with the flux limited source sub-sample indicated above. The solid line shows the theoretical prediction using expression \autoref{eq:shorthand} computed assuming a $\Lambda$CDM Cosmology \citep{2016A&A...594A..13P} and the previously measured galaxy-bias $b_{\rm L} = 1.07$ \citep{2016MNRAS.455.4301C}. The dashed line is an eye-guide for zero.}
\label{fig:resultSV}
\end{figure*}

Measured two-point angular cross-correlation functions and $\Lambda$CDM weak lensing theoretical predictions can be found in \autoref{fig:resultSV}. Measured correlation functions are found to be non-zero, compatible with $\Lambda$CDM and its amplitude evolves with the magnitude cut. The magnitude cuts imposed to guarantee uniform depth make that, for this data, no negative amplitudes are expected.

\begin{figure}
\includegraphics[width=0.5\textwidth]{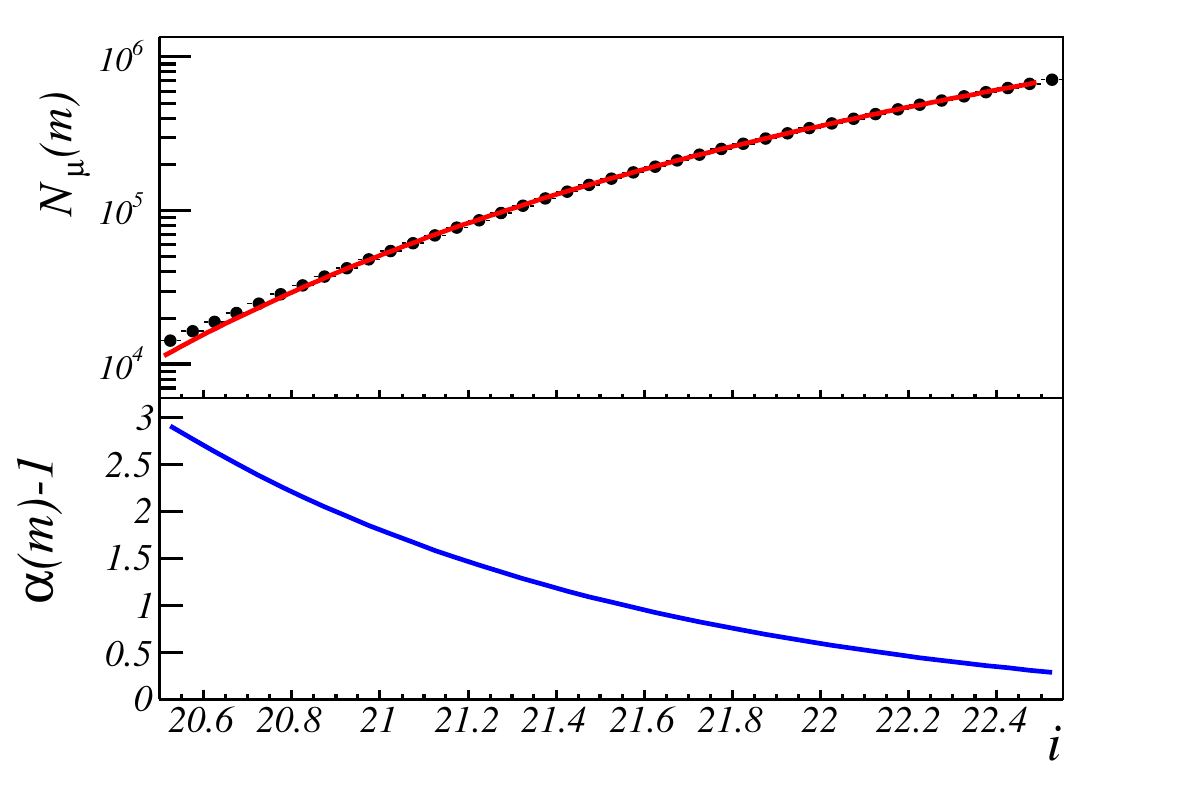}
\caption{Top panel: Dots are the measured {\it i}-band cumulative number count as a function of the {\it i}-band magnitude. Red solid line is the fit using a Schechter function (see text). Bottom panel: number count slope $\alpha-1$ measured from the fitted Schechter function of the top panel.}
\label{fig:alphai}
\end{figure}

To compare with the expected theory, \autoref{eq:4b} has been used assuming \cite{2016A&A...594A..13P} cosmological parameters. The bias of the lens sample has already been measured independently with different techniques: clustering \citep{2016MNRAS.455.4301C}, gg-lensing \citep{2016arXiv160908167P}, shear \citep{2016MNRAS.459.3203C} and CMB-lensing \citep{2016MNRAS.456.3213G}. From these values the most precise, from \cite{2016MNRAS.455.4301C}, is selected ($b_{\rm L}=1.07\pm0.08$) and is assumed to be a constant scale-independent parameter. The number count slope parameter $\alpha_{\rm S}$ is computed by fitting the cumulative number count of the sample S to a Schechter function \citep{1976ApJ...203..297S} on the range of interest
\begin{equation}
N_\mu(m) = A\left[10^{0.4(m-m_*)}\right]^\beta\times\exp\left[-10^{0.4(m-m*)}\right],
\label{eq:sch}
\end{equation}
where $A,m_*,\beta$ are the free parameters of the fit. Then $\alpha_{\rm S}(m)-1$ is computed by applying \autoref{eq:alpha}, where $m_{\rm j}$ is the magnitude limit of the $\rm S_j$ sub-sample on the considered band. In \autoref{fig:alphai} the fit and the number count slope parameter for the I sample are shown.

\begin{table}
\begin{center}\begin{tabular}{ c | c c | c c }
Sample & $\log_{10}\mathcal{B}$ & $\chi^2/ndof$  & $\log_{10}\mathcal{B}$ & $\chi^2/ndof$ \\
\hline
$r<21.0$ &-0.3 & 1.9/6 & & \\
$r<21.5$ & 0.8 & 0.8/6 & & \\
$r<22.0$ & 2.0 & 6.6/6 & 3.9 & 21.6/30\\
$r<22.5$ & 2.3 & 7.0/6 & & \\
$r<23.0$ & 1.1 & 4.2/6 & & \\
 \hline
$i<20.5$ & 0.2 & 0.9/6 & & \\
$i<21.0$ & 2.1 & 2.0/6 & & \\
$i<21.5$ & 2.5 & 4.5/6 & 3.5 & 24.2/30\\
$i<22.0$ & 1.0 & 1.7/6 & & \\
$i<22.5$ & 0.0 & 1.5/6 & & \\
 \hline
$z<20.0$ & -0.4& 2.6/6 & & \\
$z<20.5$ & 2.3 & 2.6/6 & & \\
$z<21.0$ & 2.6 & 8.8/6 & 3.9 & 37.9/30\\
$z<21.5$ & 0.9 & 3.5/6 & & \\
$z<22.0$ & 0.5 & 2.1/6 & & \\

\end{tabular}\end{center}
\caption{Significance of the detection of a magnification signal without weights. Significances determined by the logarithm of the Bayes factor and $\chi^2$ values of the expected theoretical signal are shown for each individual function as well for the combination of the five functions within a band (right).}
\label{tab:significance}
\end{table}

\begin{table}
\begin{center}\begin{tabular}{c | c c }
Sample & $\log_{10}\mathcal{B}$ & $\chi^2/ndof$  \\
\hline
$r<23.0$ & 3.2 & 3.2/6\\
$i < 22.5$ & 2.1&2.1/6\\
$z<22.0$ & 2.3&2.3/6\\
\end{tabular}\end{center}
\caption{Significance of the detection of a magnification signal with weights. Results are shown for the faintest sample.}
\label{tab:significance_weighted}
\end{table}

A goodness of fit test of the measured two-point angular cross-correlation function respect to the theoretical predictions for each band is performed combining the five correlations functions within each band:
\begin{eqnarray}
&\chi^2_{\rm Planck} =\sum\limits_{\eta\nu ij}[\tilde \omega_{\rm LS_i}(\theta_\eta)-\omega_{\rm LS_i}(\theta_\eta)]\\
&C^{-1}(\omega_{\rm LS_i}(\theta_\eta);\omega_{\rm LS_j}(\theta_\nu))[\tilde \omega_{\rm LS_j}(\theta_\nu)-\omega_{\rm LS_j}(\theta_\nu)],
\end{eqnarray}
where $\tilde\omega,\omega$ are the measured and theoretical cross-correlation functions respectively. Goodness of fit tests are also made testing the hypothesis of absence of magnification:
\begin{eqnarray}
&\chi^2_{\rm zero}=\\
&\sum\limits_{\eta\nu ij}\tilde\omega_{\rm LS_i}(\theta_\eta)C^{-1}(\omega_{\rm LS_i}(\theta_\eta);\omega_{\rm LS_j}(\theta_\nu))\tilde\omega_{\rm LS_j}(\theta_\nu)\nonumber.
\end{eqnarray}
The $\chi^2$ values of the individual correlation functions as well as the combination of the five correlation functions within each band can be seen in \autoref{tab:significance} showing good agreement with the theoretical predictions described in \autoref{sec:mag_theory}. To test which hypothesis is favored, the Bayes factor is used:
\begin{equation}
\mathcal{B} = \frac{P(M|\Theta)}{P(Z|\Theta)} = \frac{P(\Theta |M)}{P(\Theta|Z)}\frac{P(M)}{P(Z)},
\end{equation}
where 
\begin{equation}
P(M|\Theta) = e^{-\chi^2_{\rm Planck}/2}
\end{equation}
and
\begin{equation}
P(Z|\Theta) = e^{-\chi^2_{\rm zero}/2}.
\end{equation}
The assumed prior sets detection and non-detection of magnification to be equally probable: $P(M) = P(Z)$. Bayes factors are computed for each function individually as well as for each band using the full covariance.

The significance for each individual correlation function (see \autoref{tab:significance}) has a strong dependence on the considered magnitude limit of the sub-sample. To compute the significance of the detection for each band, the five correlation functions within each and the full covariance are used. One covariance matrix (see \autoref{fig:cov_matrix} for the {\itshape i}-band matrix) per each band is computed taking into account the full set of correlations. The logarithm of the Bayes factor can be found in \autoref{tab:significance}, being all above 2, allowing to claim that magnification has been detected \citep{10.2307/2291091}.
 
\begin{figure}
\includegraphics[width=0.5\textwidth]{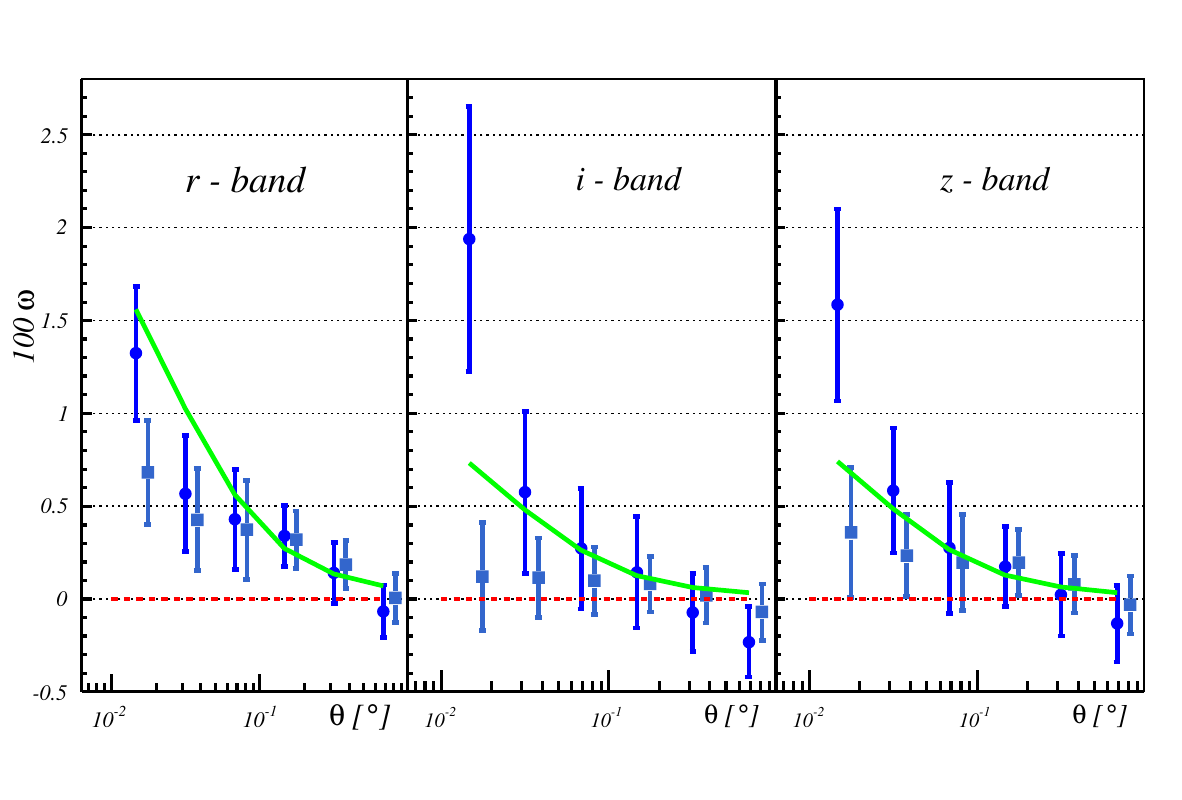}
\caption{Measured two-point angular cross-correlation functions for the samples $r<23.0$, $i < 22.5$ and $z < 22.0$ left to right respectively. Dots use the optimal weighting \citep{0004-637X-633-2-589}, where each galaxy is weighted by its corresponding $\alpha_S(m)-1$ value, whereas squares are not weighted. Green line is the theoretical prediction. Red dashed line is an eye-guide for zero.}
\label{fig:reweight}
\end{figure}

 A usual approach to enhance the signal-to-noise ratio, is to define a unique source sample, weight each source galaxy with its corresponding $\alpha_S(m)-1$ value \citep{2003A&A...403..817M} and compute the two-point angular cross-correlation function. This weighting procedure is used at the samples $r < 23.0$, $i < 22.5$ and $z < 22.0$. These correlation functions can be seen in \autoref{fig:reweight} with a comparison with the theoretical prediction and the correlation functions of the same sample computed without weighting. Significances of these measurement can be seen at \autoref{tab:significance_weighted} finding that the weighting approach provides an enhancement of the significance when compared with an unique sample. However, when the full set of correlation functions and their covariances are used, the results are similar since the same amount of data and information is used, that is, the number-count slope.
 
\begin{figure}
\includegraphics[width=0.5\textwidth]{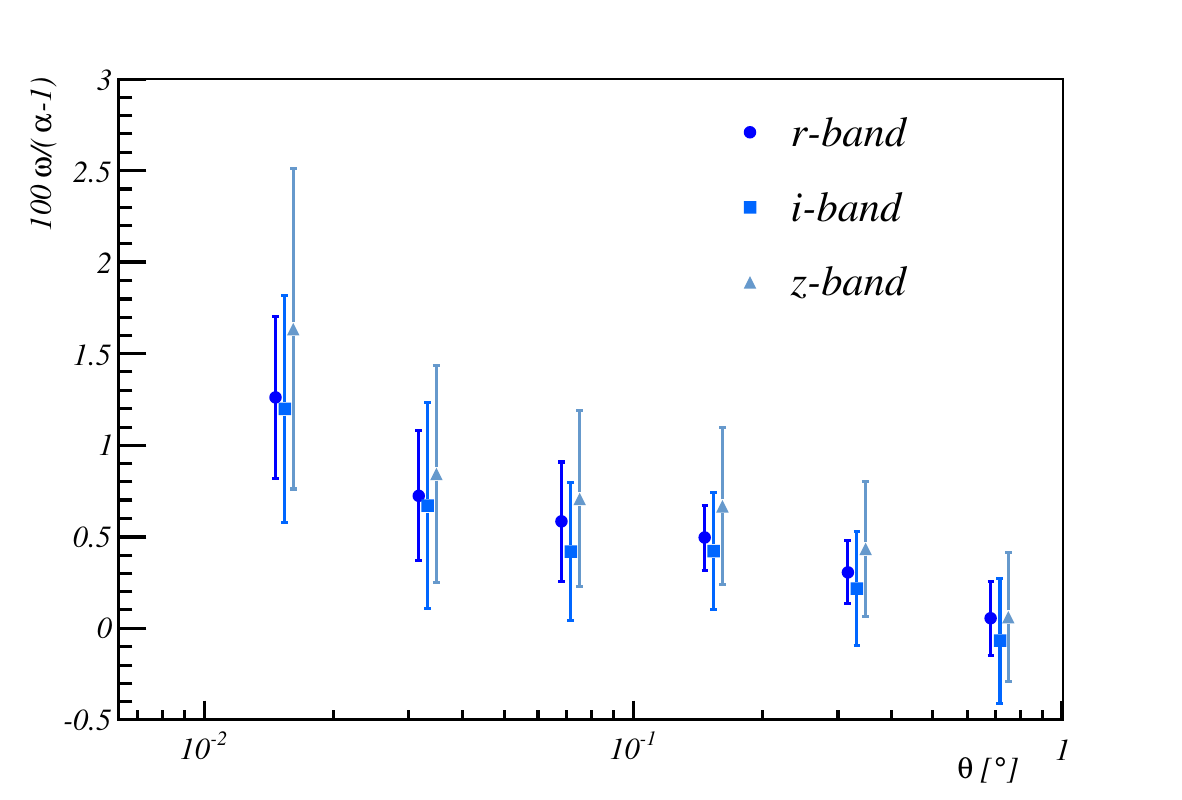}
\caption{Example of the achromaticity of the measured signal. Here are shown the measured two-point angular cross-correlation functions for $r<22.5$, $i<22.0$ and $z<21.5$ divided by their corresponding $\alpha-1$.}
\label{fig:achrom}
\end{figure}

 Finally, in order to test that the signal is achromatic, the measured two-point angular cross-correlation functions for each band, normalized by its $\alpha_S(m)-1$ are compared. All cross-correlation functions fluctuate within $1\sigma$ errors (see \autoref{fig:achrom} for an example) demonstrating that the measured convergence field does not depend on the considered band.

\subsection{Systematic errors}
\label{sec:sys}

In this section, the impact of potential sources of systematic errors on the measured two-point angular cross-correlation function is investigated and how they are taken into account in the measurement is described.

\subsubsection{Number count slope $\alpha$}
    
\begin{figure}
\includegraphics[width=0.5\textwidth]{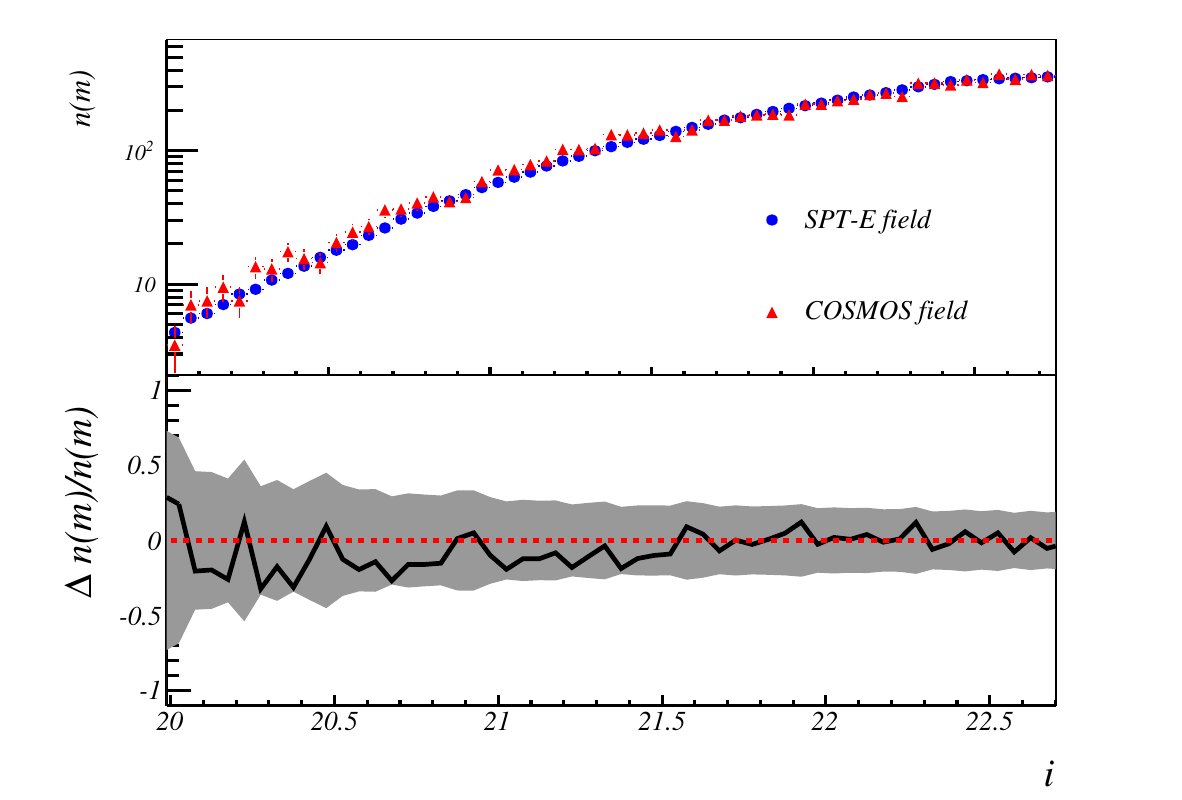}
\caption{Upper panel: Comparison of the magnitude distribution for the SPT-E and the COSMOS fields. Both histograms are normalized by their respective area. Lower panel: Relative difference between the magnitude distribution of the COSMOS and the SPT-E fields. The shaded region shows the $1\sigma$ confidence interval computed from shot-noise.}
\label{fig:ndmSN}
\end{figure}
	When comparing the measured two point angular cross-correlation functions with the theoretical prediction via \autoref{eq:shorthand} for a given set of cosmological parameters, $\alpha(m)$ is determined by fitting the cumulative number count distribution to \autoref{eq:sch} and then using \autoref{eq:alpha}. To compute the possible impact of the uncertainty of this fit on the comparison with theory, a marginalisation over all the parameters of the fit ($A,m_*,\beta$) is made.
	
Parameters are randomly sampled with a Gaussian distribution centred on the value given by the fit to the cumulative number count and with a standard deviation equal to the $1\sigma$ errors of the fit. The value of $\alpha$ is recalculated with these randomly sampled parameters. The impact of the dispersion of the $\alpha$ values obtained is negligible compared to the size of the jackknife errors, so they are not taken into account.

In addition to the parameter determination, a possible non-completeness on the SPT-E field can modify the magnitude distribution altering the cumulative number count slope parameter \citep{2016MNRAS.455.3943H}. To estimate the possible impact of non-completeness, the measured magnitude distributions of the SPT-E field are compared with those of deeper fields measured by DES, such as the COSMOS field. Both distributions are found to be equal at the range of magnitudes considered on this analysis (see \autoref{fig:ndmSN} for an example in the $i$-band).

\subsubsection{Object obscuration}

\cite{0004-637X-801-2-73} studied whether moderately bright objects in crowded environments produce a decrease in the detection probability of nearby fainter objects at scales $\theta\lesssim10\ $arcsec. However, such scales are well below those considered in this analysis ($\theta>36$ arcsec) and therefore this effect is ignored.

\subsubsection{Stellar contamination}
    
	For a given choice of star-galaxy classifier, there will be a number of stars misclassified as galaxies, so the observed two-point angular cross-correlation function $\omega_O(\theta)$ must be corrected by the presence of any fake signal induced by stars (see \cref{sec:starscorrection}):
	\begin{equation}
	\omega_{\rm LS_j} = \frac{\omega_{\rm O}(\theta)-\lambda_{\rm L}\omega_{\rm *S_j}(\theta)-\lambda_{\rm S_j}\omega_{\rm L*}(\theta)}{1-\lambda_{\rm L}-\lambda_{\rm S_j}},
	\label{eq:starcorrection}
	\end{equation}
	where $\omega_{\rm LS_j}$ is the corrected galaxy cross-correlation function, $\omega_{\rm L*}$ is the cross-correlation function of the true galaxy lenses with the stars misclassified as galaxies in the source sample, $\omega_{\rm *S_j}$ is the cross correlation of the stars misclassified as galaxies in the lenses with the true source galaxies and $\lambda_{\rm L},\lambda_{\rm S_j}$ are the fraction of stars in the lens and in the source samples respectively.
	Assuming that the misclassification of stars is spatially random and is a representative sample of the spatial distribution of the population classified as stars and that the fraction of misclassified stars is small, the functions $\omega_{\rm L*},\omega_{\rm *S_j}$ are estimated from the cross-correlation of the galaxy population and the stellar population in the corresponding redshift bin.

    \begin{figure}
	\includegraphics[width=0.5\textwidth]{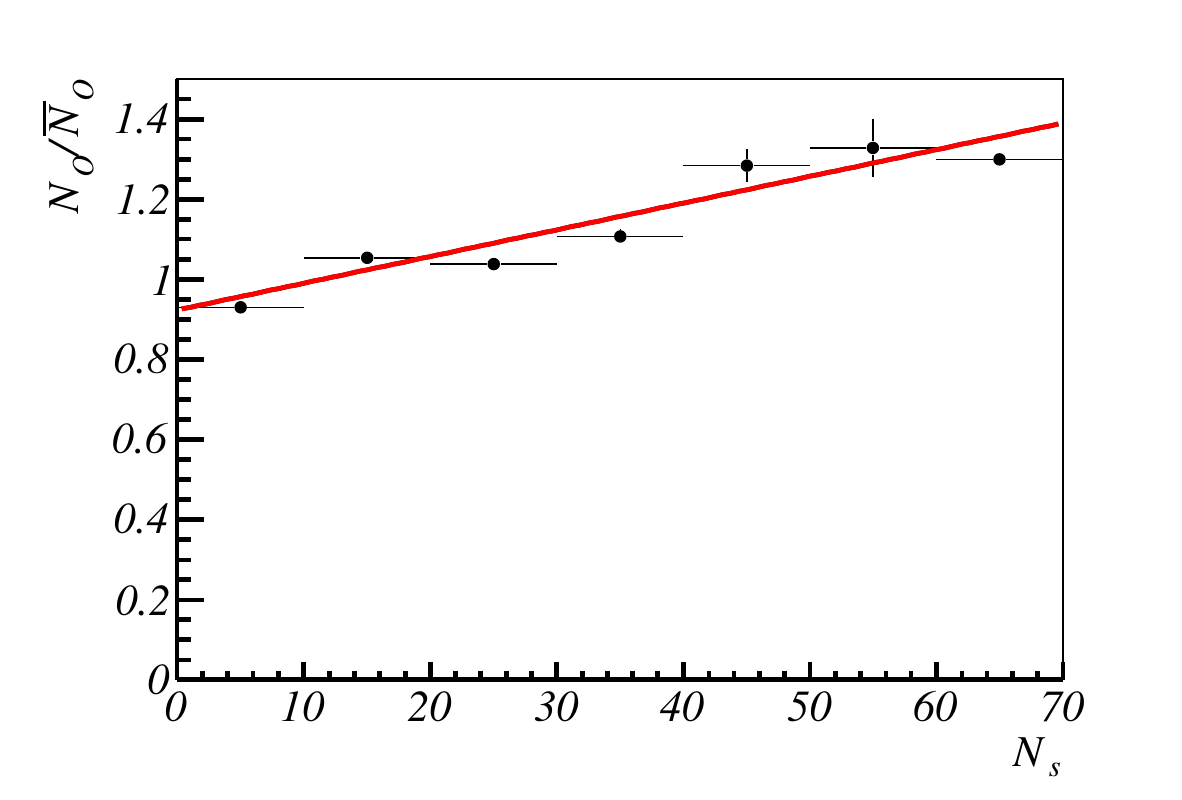}
    \caption{Determination of the purity of the lens sample. For each $N_{\rm nside}=512$ {\scshape HEALPix}-pixel, the number of objects classified as galaxies divided by the average number of galaxies per pixel is plotted as a function of the  number of objects classified as stars. Black dots are the measured data. Red line is the linear fit to the data. The intercept of the line with the Y-axis is the estimated purity of the sample.}
    \label{fig:purity}
	\end{figure}
    \begin{figure}
\includegraphics[width=0.5\textwidth]{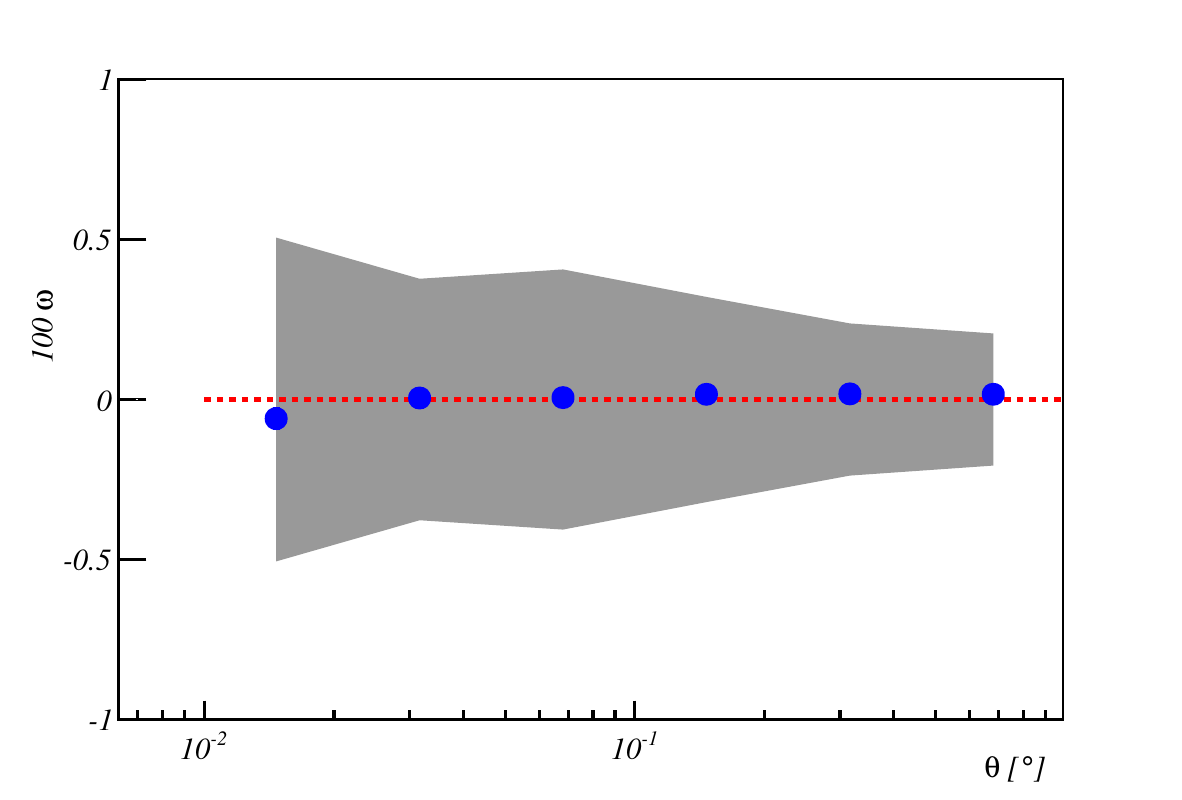}
\caption{Correction by stellar contamination on the $i<21.5$ sample. Blue dots are the correction and shaded area is the $1\sigma$ confidence interval of the measured cross-correlations of the magnification signal. Red dashed line is and eye-guide for zero.}
\label{fig:correction}
\end{figure}
 
	Following a similar approach to \cite{2012MNRAS.424..564R}, if the latter is true and the misclassified stars trace the global population of stars, for a given patch of the sky the number of objects classified as galaxies $N_{\rm O}$ must be the average number of true galaxies $\bar N_{\rm g}$ plus a quantity proportional to the number of stars on that given pixel,
	\begin{equation}
	N_{\rm O} = \bar N_{\rm g}+\tilde\gamma N_{\rm s}.
	\end{equation}
	Dividing by the average number of objects marked as galaxies $\bar N_{\rm O}$,
	\begin{equation}
	\frac{N_{\rm O}}{\bar N_{\rm O}} = p+\gamma N_{\rm s},
	\label{eq:purity}
	\end{equation}
	where $p=\bar N_{\rm g}/\bar N_{\rm O}$ is the purity of the sample, that is, $\lambda = 1-p$.
	
	In order to estimate the purity of the galaxy sample with this method, an $N_{\rm side}=512$ {\scshape HEALPix} pixelation is made and for each pixel $N_{\rm O}/\bar N_{\rm O}$ and $N_{\rm s}$ is computed. Then, a fit to \autoref{eq:purity} is made determining a purity of 94 per cent for the lens sample and about 98 per cent for the source sample depending on the considered band (see \autoref{fig:purity} for an example). With this purity, the correction due to stellar contamination given by \autoref{eq:starcorrection} is found to be one order of magnitude smaller than the statistical errors (see \autoref{fig:correction} for the $i$-band correction), so stellar contamination is not taken into account in the analysis. Nevertheless, on future analysis with more galaxies and area this may be important.
    Note that the objects labeled as stars by our star-galaxy classifier would be a combination of stars and galaxies thus these calculations are an upper bound to stellar contamination.

	\subsubsection{Survey observing conditions}
    \label{sec:balrog}
    Observing conditions are not constant during the survey, leading to spatial dependencies across the DES-SV footprint \citep{2015arXiv150705647L} that may affect the observed cross-correlation function, such as seeing variations, air-mass, sky-brightness or exposure time \citep{2015MNRAS.454.3121M}. To trace these spatial variations, the catalog produced by the Monte Carlo sampling code {\scshape Balrog} has been used as random sample \citep{2016MNRAS.457..786S}. It is important to remark that {\scshape Balrog} catalogs are produced with the same pipeline as DES-SV data, allowing one to trace subtle effects such as patchiness on the zeropoints, deblending and possible magnitude errors due to a wrong sky subtraction close to bright objects.
    
     The use of Monte Carlo sampling methods provides a new approach to mitigate systematic effects complementary to methods that cross-correlate the galaxy-positions with the maps of the survey observing conditions \citep{2012MNRAS.424..564R,2012ApJ...761...14H,2015MNRAS.454.3121M} or involve masking the regions of the sky with worst values of the observing conditions \citep{2016MNRAS.455.4301C}. The amount of sky to be masked in order to mitigate the systematic effects on the correlation functions, is freely decided based on the impact on the correlation function, which may lead to a biassed measurement. On the other hand, the approach involving cross-correlations may lead to an overcorrection effect since the different maps of the observing conditions are, in general, correlated in a complicated manner \citep{2016MNRAS.456.2095E}. This new Monte Carlo technique to sample the selection function of the survey given by {\scshape Balrog}, has the advantage that takes into account the correlation of the different observing conditions maps as well as provides an objective criteria to mitigate systematic errors on the correlation function for a given sample, avoiding biassed measurements. In addition, the use of {\scshape Balrog} has the potential to allow us in the future to exploit the full depth of the survey \citep{2016MNRAS.457..786S}.

\subsubsection{Dust extinction}
\label{sec:dustext}

\begin{figure}
\includegraphics[width=0.5\textwidth]{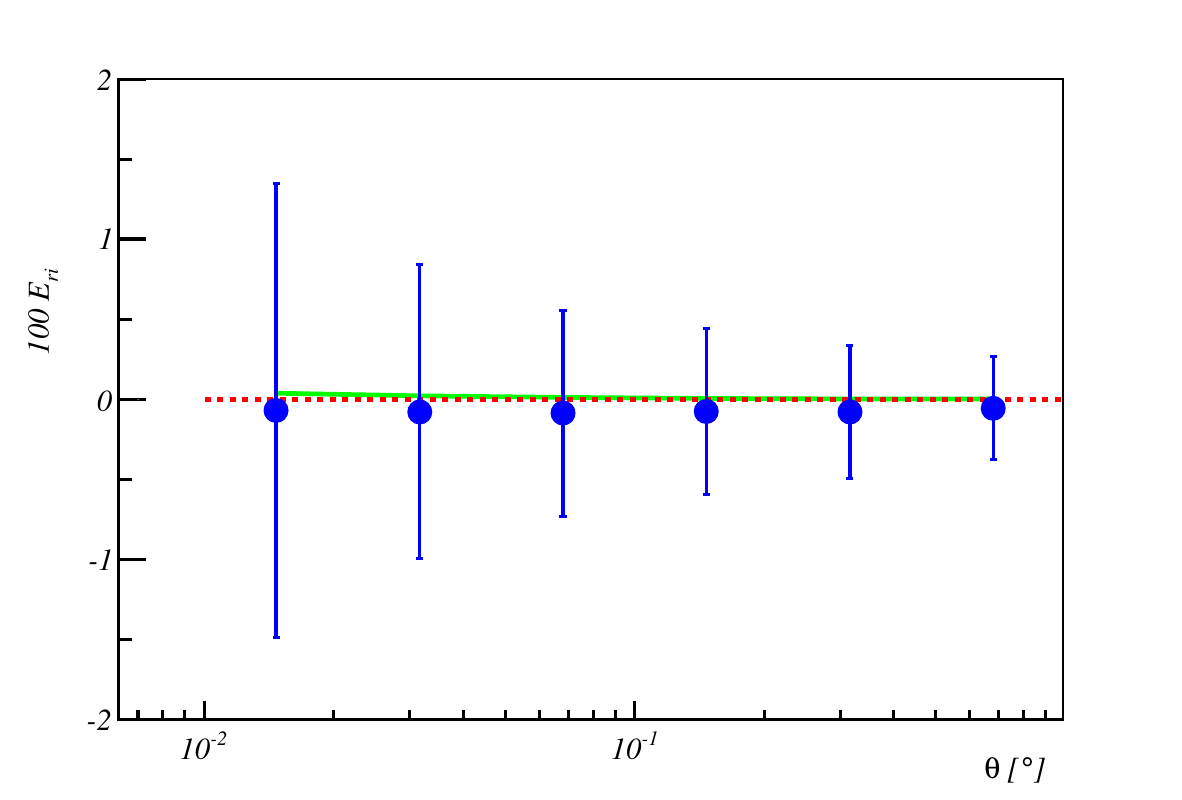}
\caption{Blue dots: colour-density cross-correlation functions measured on SV data for the {\it r} and {\it i} bands (sample $i<21.5$). Green solid line is the expected value from \autoref{eq:colorexcess}. Red dashed line is an eye-guide for zero.}
\label{fig:colorexcess}
\end{figure}

\begin{figure}
\includegraphics[width=0.5\textwidth]{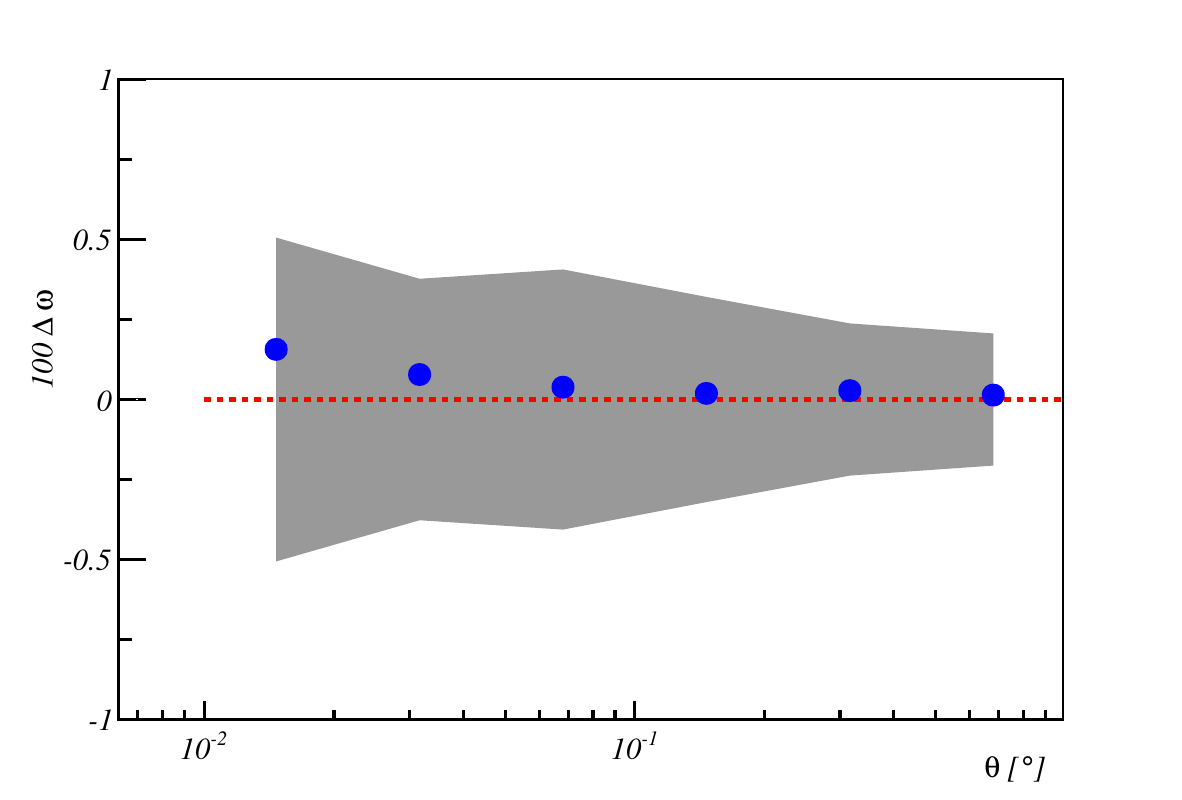}
\caption{Impact of dust on the number count from MICE (case $i<21.5$). Shade is the $1\sigma$ confidence interval. Blue dots are the number count differences between the case with and the case without the simulated dust profile. Red dashed line is an eye-guide for zero.}
\label{fig:micedust}
\end{figure}
The possible presence of dust in the lenses may modify the observed magnitude in addition to the magnitude shift due to magnification \citep{2010MNRAS.405.1025M}. The change in magnitude ($\delta m$) on the $p$-band may be written as
\begin{equation}
\delta m_p = -2.5\log\mu+\frac{2.5}{\ln10}\tau_p,
\end{equation}
where  $\mu\simeq1+2\kappa$ is the change in magnitude due to magnification and $\tau_k$ is the optical depth due to dust extinction. Whereas magnification is achromatic, dust extinction induces a band-dependent magnitude change. Taking this into account, the colour-excess for bands $p,q$\footnote{In this section $p,q$ stand for a generic index label while $V$ stands for the $V$ band of the $UBV$ system.} is defined as
\begin{equation}
E_{pq} = \delta m_p-\delta m_q=1.08[\tau_p-\tau_q].
\end{equation}
Define the colour-density cross-correlation as \citep{2010MNRAS.405.1025M}
\begin{equation}
\langle \delta_{\rm g}E_{pq}\rangle(\theta) = 1.09[\tau_p(\theta)-\tau_q(\theta)],
\end{equation}
where $\delta_{\rm g}$ is the density contrast of the lenses and $E_{pq}$ is the colour-excess of the sources; from the measurements by \cite{2010MNRAS.405.1025M} it can be parametrized as
\begin{equation}
\langle\delta_{\rm g}E_{pq}\rangle(\theta) = 1.09\tau_V\left[\frac{\lambda_V}{\lambda_p}-\frac{\lambda_V}{\lambda_q}\right]\left(\frac{\theta}{1'}\right)^{-0.8},
\label{eq:colorexcess}
\end{equation}
with $\tau_V=2.3\times10^{-3}$ the optical depth at the {\it V}-band  and $\lambda_V,\lambda_p,\lambda_q$ the average wavelengths of the $V$, $p$ and $q$ bands respectively. With this parametrization, the impact of dust extinction is negligible at the scales considered on this analysis. As it can be seen in \autoref{fig:colorexcess}, colour-density cross-correlation functions are compatible with \autoref{eq:colorexcess} as well as with zero.

In addition, the impact of a dust profile has been simulated as described in \autoref{eq:colorexcess} with the MICE simulation (\autoref{sec:simulation}). To do so, for each galaxy belonging to the source sample a magnitude shift is induced
\begin{equation}
m_d = m_\mu +1.09\tau_V\frac{\lambda_V}{\lambda}\sum\limits_{l}\left(\frac{\theta_l}{1'}\right)^{-0.8}.
\end{equation}
Here $\theta_l$ is the angular separation of the source-galaxy and the $l$-th lens galaxy and the summation is over all the galaxies of the lens sample. In \autoref{fig:micedust} the difference between the two-point angular cross-correlation with and without the dust can be seen to be less than the statistical errors. It can be deduced that dust has no impact on the angular scales considered on this work.

Since the parametrization used here only applies to a sample similar to the one used at \cite{2010MNRAS.405.1025M}, statements about dust constrains are limited. Nevertheless this does not change the fact that no chromatic effects are detected.

\subsubsection{Photometric redshifts}

\begin{figure}
\includegraphics[width=0.5\textwidth]{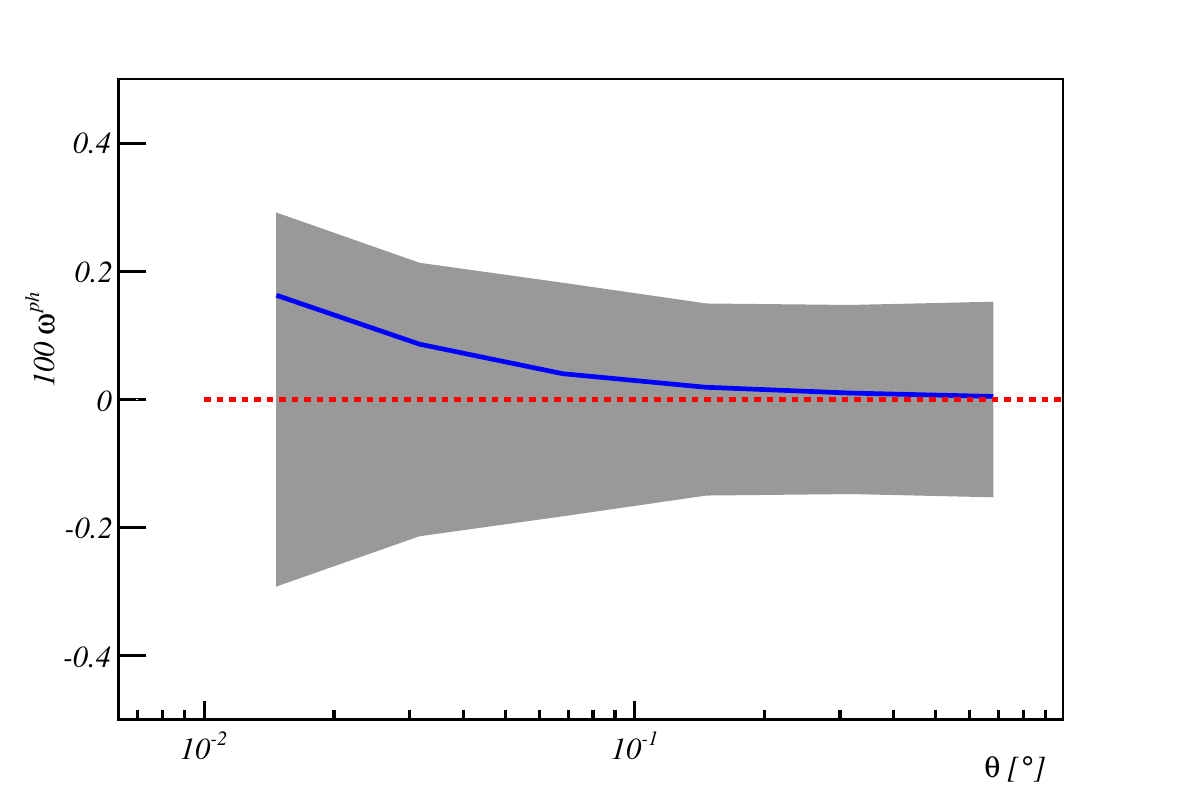}
\caption{Comparison of $1\sigma$ jackknife errors of the measured correlation function (grey shade) with the expected signal induced by the photo-z migration between the lens and the source sample (case $i<22.5$) computed theoretically with the stacking of the pdfs for the $i$-band (blue line).}
\label{fig:theophotoz}
\end{figure}

A general study of photo-z performance in DES-SV can be found in \cite{2014MNRAS.445.1482S}. A comprehensive study of the photo-z performance and its implications for weak lensing can be found in \cite{PhysRevD.94.042005}. Both studies are followed in this analysis.
	Conservative photo-z cuts are made in order to minimize the migration between lens and source samples. Nevertheless, catastrophic outliers in the photo-z determination can bias the measurement of  $\kappa$ \citep{2010MNRAS.401.1399B}. Thus, the tails of the probability density functions (pdfs) of the photo-z code are a crucial systematic to test.
	As mentioned in \autoref{sec:mag_theory}, in addition to the magnification signal, galaxy migration due to a wrong photo-z assignment between lens and source samples may induce a non-zero cross-correlation signal due to the physical signal coming from the clustering of objects in the same redshift bin. As a first approach, estimation of the expected signal induced by photo-z migration ($\omega^{ph}$) is computed with \autoref{eq:4a}:
\begin{equation}
\omega_{\rm LS_j}^{\rm ph}(\theta) = \int\limits_0^\infty dz\int\limits_0^\infty dz' \phi_{\rm L}(z)\phi_{\rm S_j}(z')\xi(\theta;z,z'),
\label{eq:phth}
\end{equation}
where $\xi(\theta;z,z')$ is the 3D correlation-function and $\phi_{\rm L},\phi_{\rm S_j}$ are the redshift distribution of the lens (L) sample and the source sample ($\rm S_j$) estimated from the stacking of the pdfs given by TPZ. \autoref{fig:theophotoz} compares the measured two-point angular cross-correlation and the expected signal induced by photo-z can be seen for the \textit{I} sample. The signal induced by photo-z is found to be smaller than the statistical errors. Note that this method relies on an assumed cosmology and bias model, and therefore should be considered only an approximation. A more accurate calculation can be made with the help of N-body simulations.

From the overlap of the redshift distribution of both lens and source samples, it is found that the total photo-z migration between lens and source sample is $o\sim 0.6\%$ depending on the magnitude cut of the source sample. The procedure to compute this overlap is to integrate the product of the pdfs of the lens and source sample:
\begin{equation}
o = \int\limits_0^\infty dz\phi_L(z)\phi_S(z),
\end{equation}
where $\phi_L,\phi_S$ are the stacked pdfs of the lens and source sample respectively. Since TPZ provides an individual pdf for each galaxy, the stacked pdf of a given sample is computed by adding all the individual pdfs of the galaxies that belong to that sample (see \cite{2016MNRAS.459.1293A} for a study of clustering with stacked pdfs).

\begin{figure}
\includegraphics[width=0.5\textwidth]{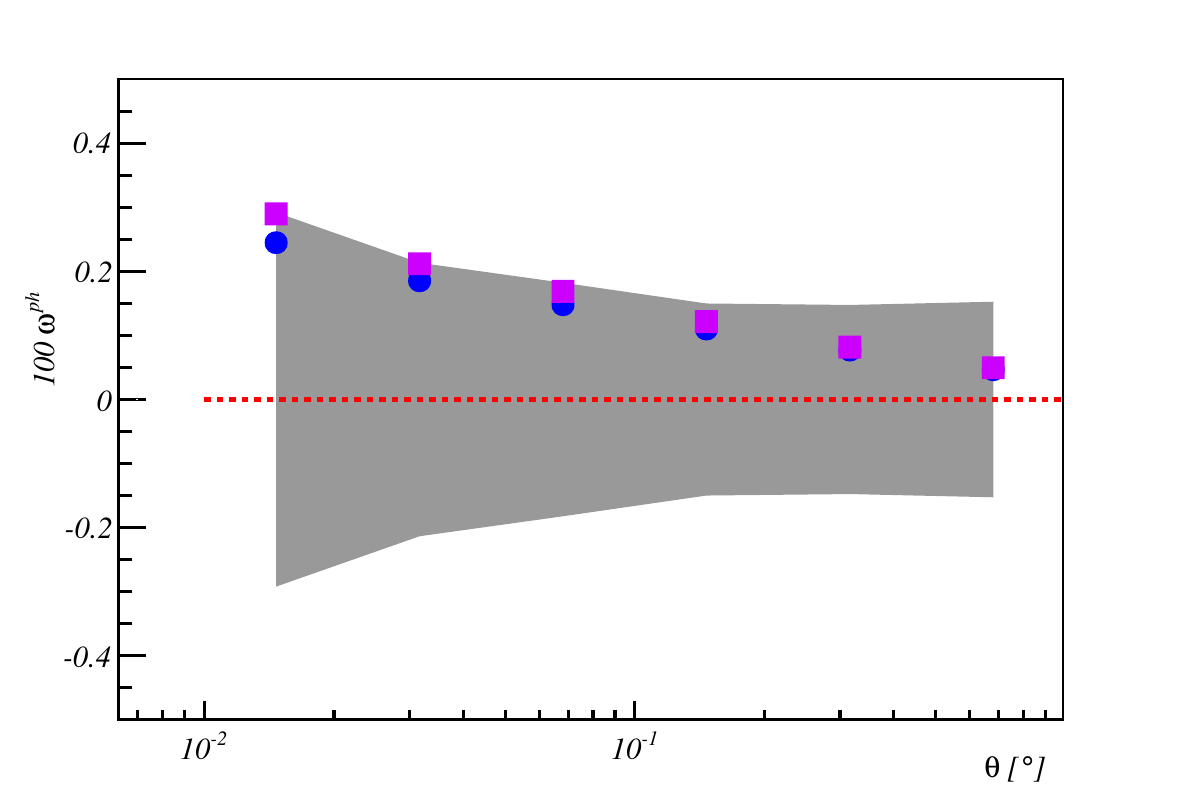}
\caption{Estimation of the signal induced by migration of selected fractions of MICE un-lensed galaxies between the lens and the source sample (case $i<22.5$). Shaded area is the $1\sigma$ confidence interval for the measured number count cross-correlations. Dots correspond to a contamination fraction of 0.9 per cent. Squares correspond to a 2 per cent. Dashed line is an eye-guide for zero.}
\label{fig:photozcontamination}
\end{figure}

To estimate the maximum allowed photo-z migration between the lens and the source sample, the MICE simulation (\autoref{sec:simulation}) with the un-lensed coordinates and magnitudes is used. Galaxies are randomly sampled on the lens redshift bin and then placed on the source redshift bin. Conversely, galaxies on the source redshift bin are randomly sampled and placed on the lens redshift bin. For a given lens or source sample, the number of galaxies introduced from the other redshift bin is chosen to be 0.1, 0.3, 0.5, 0.7, 0.9 and 2 per cent of the galaxies. Then, the two-point angular cross-correlation is computed for each case. The difference of the correlation functions measured at the simulation with induced migration between lens and source sample and the original used in \autoref{sec:simulation} is the signal induced by photo-z migration. The signal induced by photo-z for the cases with 0.9 and 2 per cent computed with this method can be seen at \autoref{fig:photozcontamination}. It is found that  at 0.9 per cent of contamination, the induced signal due to photo-z migration is comparable to the error in the correlation functions. This upper limit is greater than the estimated photo-z migration, demonstrating that the effect of photo-z migration is negligible. Photo-z migration has a larger impact on the brightest samples. Nevertheless, since the errors of the correlation functions of these samples are shot-noise dominated, the tightest constraints on photo-z migration are imposed by the faintest samples. With a larger data sample this statement will no longer be true.

Photo-z induced correlation functions that mimic magnification may affect the measured significance. Thus, Bayes factor is recomputed with two new hypothesis, the measured signal is a combination of magnification and photo-z ($M+Ph$) or the measured signal is only photo-z ($Ph$):
\begin{equation}
\mathcal{B} = \frac{P(M+Ph|\Theta)}{P(Ph|\Theta)} = \frac{P(\Theta|M+Ph)}{P(\Theta|Ph)},
\end{equation}
where
\begin{equation}
P(\Theta|M+Ph) = e^{-\chi^2_{\rm Planck+Ph}/2}
\end{equation}
and
\begin{equation}
P(\Theta|Ph) = e^{-\chi^2_{\rm Ph}/2}.
\end{equation}
To compute $\chi^2_{\rm Planck+Ph}$ and $\chi^2_{\rm Ph}$ it has been assumed that the expected theory is given by $\omega_{\rm LS_j}(\theta)+\omega_{\rm LS_j}^{\rm ph}(\theta)$ and $\omega_{\rm LS_j}^{\rm ph}$ respectively, where $\omega_{\rm LS_j}^{\rm ph}$ is the expected signal induced by photo-z computed using \autoref{eq:phth}. The significances recomputed using these two new hypothesis for the {\it r}, {\it i} and {\it z} bands are $\log_{10}\mathcal{B}=2.5, 4.0, 3.5$ respectively. Thus, it can be concluded that photo-z migration does not change the conclusions.

All previous calculations were based on the assumption that the pdfs are a reliable description of the true redshift distribution. This statement has been validated by previous works (\cite{2014MNRAS.445.1482S}; \cite{PhysRevD.94.042005}).  Redshift distributions predicted by TPZ are found to be representative of those given by the spectroscopic sample. Nevertheless, this statement has limitations --but is good enough for SV data-- and a more accurate description of the real redshift distribution of the full sample will be measured with methodologies involving clustering-based estimators \citep{2008ApJ...684...88N,2010ApJ...721..456M,2013arXiv1303.4722M,2016MNRAS.462.1683S} when the size of the data sample grows.

\begin{figure}
\includegraphics[width=0.5\textwidth]{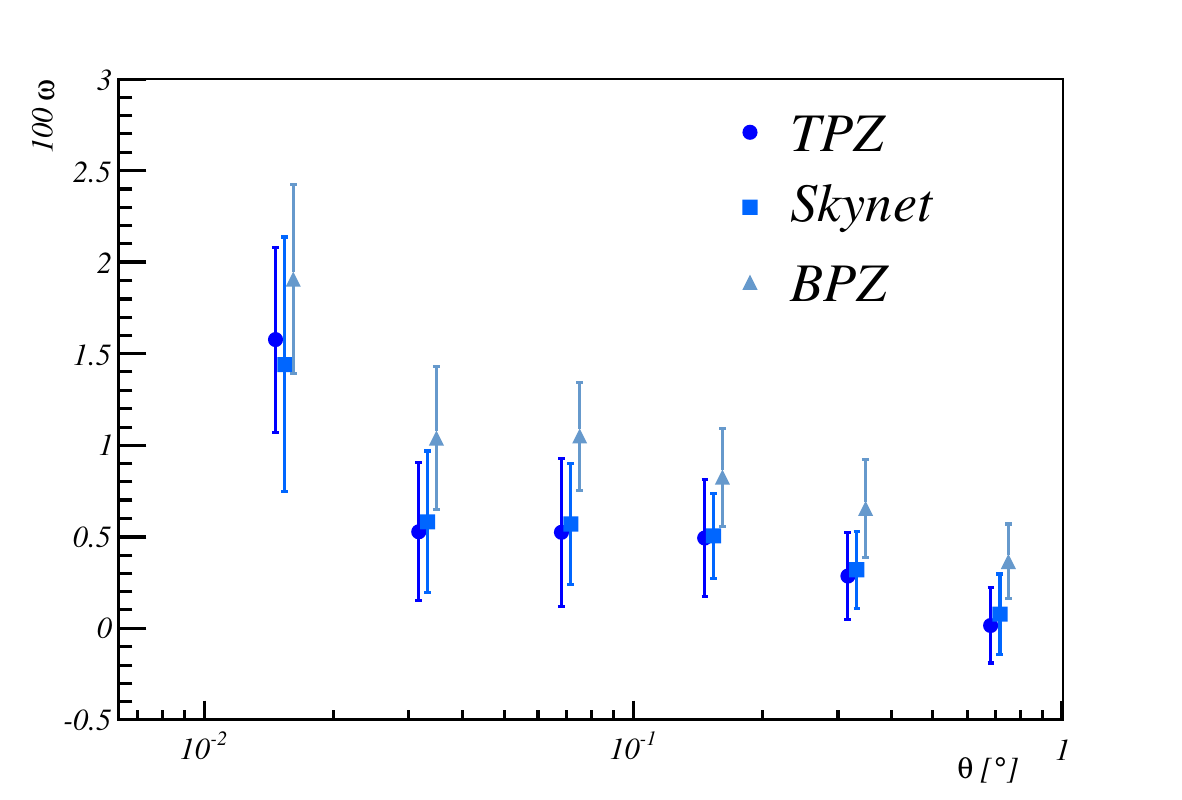}
\caption{Comparison of the measured two-point angular cross-correlation functions corresponding to the sample $i<21.5$ measured with the Landy-Szalay estimator using TPZ, Skynet and BPZ. Triangles and squares are displaced at the horizontal axis for clarity.}
\label{fig:bpzskynet}
\end{figure}

    Finally, to demonstrate that the measured signal is independent of the photo-z technique employed to estimate the redshift, the two-point angular cross-correlation functions  are measured using other estimators for photo-z, and have a performance similar to TPZ \citep{2014MNRAS.445.1482S}. A neural network, Skynet  \citep{2014MNRAS.441.1741G}, and a template based approach, Bayesian Photo-Z (BPZ; \cite{2000ApJ...536..571B}) have been used. \autoref{fig:bpzskynet} compares the cross-correlations computed with the three codes for the $i$-band, showing them to be within $1\sigma$ errors.
    
\section{Conclusions}
\label{sec:conclusions}

In this paper weak lensing magnification of number count has been detected with the Dark Energy Survey, on each of the {\it r}, {\it i}, {\it z} photometric bands. The measured magnification signal agrees with theoretical predictions using a $\Lambda$CDM model with \cite{2016A&A...594A..13P} best-fit parameters.

This magnification measurement has been made using all measured galaxies, selecting them only by its photo-z. A method that makes explicit use of the full set of covariance matrices to maximize the significance has been used. The proposed method is compared with the usual weighting approach that can be found on the literature, reaching a similar level of significance. Although the methodology proposed on this work does not improve the signal-to-noise of the measurement of number-count magnification, allows a better control and estimation of systematic effects. Systematic effects due to observing conditions or photo-z have an strong correlation with the magnitudes of the galaxies. Thus, the weighted combination of galaxies leads to a non-trivial combination of correlated systematic effects. The proposed methodology analyzes independently the systematic effects of each set of galaxies combining afterwards the measured number-count magnification signal already corrected.

Systematic effects have been studied in detail not only using the data itself, but also supported with the N-body simulation {\scshape MICE} and the {\scshape Balrog} Monte Carlo sampling method. The use of {\scshape Balrog} provides a new and powerful way to deal with systematic errors complementary to usual approaches as masking and cross-correlations as it has been stated in \autoref{sec:balrog}

The detection of magnification has been made only with 3 per cent of the final planned area for DES and half of the available maximum depth. This demonstrates that magnification measurements are feasible in the Dark Energy Survey and can provide an useful complement to the survey's main goal on future data releases covering wider areas of the sky.

Future work will include the analysis of DES observations in much wider area, where some of the systematic issues not significant here such as stellar contamination and the accurate determination of the number count slope parameter, may not be negligible. These analyses will include measurements of cosmological parameters --by themselves or in combination with other weak lensing measurements \citep{2010MNRAS.401.2093V}--, but also the other two effects of magnification: the observed magnitude shift \citep{2010MNRAS.405.1025M} and the increase in the observed size \citep{2041-8205-780-2-L16}.

\section*{Acknowledgements}
\label{sec:acknowledgements}  
We are grateful for the extraordinary contributions of our CTIO colleagues and the DECam Construction, Commissioning and Science Verification teams in achieving the excellent instrument and telescope conditions that have made this work possible.  The success of this project also relies critically on the expertise and dedication of the DES Data Management group.

Funding for the DES Projects has been provided by the U.S. Department of Energy, the U.S. National Science Foundation, the Ministry of Science and Education of Spain, the Science and Technology Facilities Council of the United Kingdom, the Higher Education Funding Council for England, the National Center for Supercomputing Applications at the University of Illinois at Urbana-Champaign, the Kavli Institute of Cosmological Physics at the University of Chicago, the Center for Cosmology and Astro-Particle Physics at the Ohio State University, the Mitchell Institute for Fundamental Physics and Astronomy at Texas A\&M University, Financiadora de Estudos e Projetos, Funda{\c c}{\~a}o Carlos Chagas Filho de Amparo {\`a} Pesquisa do Estado do Rio de Janeiro, Conselho Nacional de Desenvolvimento Cient{\'i}fico e Tecnol{\'o}gico and the Minist{\'e}rio da Ci{\^e}ncia, Tecnologia e Inova{\c c}{\~a}o, the Deutsche Forschungsgemeinschaft and the Collaborating Institutions in the Dark Energy Survey. 

The Collaborating Institutions are Argonne National Laboratory, the University of California at Santa Cruz, the University of Cambridge, Centro de Investigaciones Energ{\'e}ticas, Medioambientales y Tecnol{\'o}gicas-Madrid, the University of Chicago, University College London, the DES-Brazil Consortium, the University of Edinburgh, the Eidgen{\"o}ssische Technische Hochschule (ETH) Z{\"u}rich, Fermi National Accelerator Laboratory, the University of Illinois at Urbana-Champaign, the Institut de Ci{\`e}ncies de l'Espai (IEEC/CSIC), the Institut de F{\'i}sica d'Altes Energies, Lawrence Berkeley National Laboratory, the Ludwig-Maximilians Universit{\"a}t M{\"u}nchen and the associated Excellence Cluster Universe, the University of Michigan, the National Optical Astronomy Observatory, the University of Nottingham, The Ohio State University, the University of Pennsylvania, the University of Portsmouth, SLAC National Accelerator Laboratory, Stanford University, the University of Sussex, and Texas A\&M University.

The DES data management system is supported by the National Science Foundation under Grant Number AST-1138766.

The DES participants from Spanish institutions are partially supported by MINECO under grants AYA2012-39559, ESP2013-48274, FPA2015-68048, and Centro de Excelencia Severo Ochoa SEV-2012-0234 and Mar{\'i}a de Maeztu MDM-2015-0509. Research leading to these results has received funding from the European Research Council under the European Union¡¯s Seventh Framework Programme (FP7/2007-2013) including ERC grant agreements 240672, 291329, and 306478.

\bibliography{des_mag_sv}
\appendix
\section{Influence of Stellar Contamination on the Two-Point Angular Cross-Correlation}
\label{sec:starscorrection}
The observed density contrast of objects is given by
\begin{equation}
\delta_{\rm O}(\boldsymbol{\hat n},z_i) = \frac{N_{\rm g}(z_i)+N_*(z_i)}{\bar N_{\rm g}(z_i)+\bar N_*(z_i)}-1,
\end{equation}
where $N_{\rm g}, N_*$ are the number of galaxies on direction $\boldsymbol{\hat n}$ and redshift $z_i$ and stars respectively and $\bar N_{\rm g}, \bar N_*$ the average number of galaxies and stars over the footprint. The previous equation can be expressed as
\begin{equation}
\delta_{\rm O}(\boldsymbol{\hat n},z_i) = \frac{N_{\rm g}(z_i)+N_*(z_i)}{\bar N_{\rm g}(z_i)\left[1+\frac{\bar N_*(z_i)}{\bar N_{\rm g}(z_i)}\right]}-1.
\end{equation}
Taylor expanding the brackets one has,
\begin{equation}
\delta_{\rm O}(\boldsymbol{\hat n},z_i) = \frac{N_{\rm g}(z_i)+N_*(z_i)}{\bar N_{\rm g}(z_i)}\left[1-\frac{\bar N_*(z_i)}{\bar N_{\rm g}(z_i)}\right]-1
\end{equation}
and taking common factor $\bar N_*(z_i)/\bar N_{\rm g}(z_i)$,
\begin{eqnarray}
 &\delta_{\rm O}(z_i) = \left[\frac{N_{\rm g}(z_i)}{\bar N_{\rm g}(z_i)}-1\right]+\\
 &\frac{\bar N_*(z_i)}{\bar N_{\rm g}(z_i)}\left[\frac{N_*(z_i)}{\bar N_*(z_i)}-\frac{N_{\rm g}(z_i)}{\bar N_{\rm g}(z_i)}\right]-\frac{N_*(z_i)}{\bar N_{\rm g}(z_i)}.\nonumber
\end{eqnarray}
Assuming that $\bar N_* \ll \bar N_g$, the last term can be neglected and defining $\lambda_i = \bar N_*(z_i)/\bar N_{\rm g}(z_i)$ as the fraction of stars on the $i$-th sample,
\begin{equation}
\delta_{\rm O}(\boldsymbol{\hat n},z_i) = \delta_{\rm g}(\boldsymbol{\hat n},z_i)+\lambda_i[\delta_*(\boldsymbol{\hat n},z_i)-\delta_{\rm g}(\boldsymbol{\hat n},z_i)].
\end{equation}
Calculating the two point angular cross-correlation results finally in
\begin{equation}
\omega_{\rm O} = (1-\lambda_i-\lambda_j)\omega_{\rm gg}+\lambda_j\omega_{\rm g*}+\lambda_i\omega_{\rm *g}+\lambda_i\lambda_j\omega_{**}.
\end{equation}

\section*{AFFILIATIONS}
\label{sec:affiliations}
$^{1}$ Centro de Investigaciones Energ\'eticas, Medioambientales y Tecnol\'ogicas (CIEMAT), Madrid, Spain\\
$^{2}$ Computer Science and Mathematics Division, Oak Ridge National Laboratory, Oak Ridge, TN 37831\\
$^{3}$ Jet Propulsion Laboratory, California Institute of Technology, 4800 Oak Grove Dr., Pasadena, CA 91109, USA\\
$^{4}$ Institut de Ci\`encies de l'Espai, IEEC-CSIC, Campus UAB, Carrer de Can Magrans, s/n,  08193 Bellaterra, Barcelona, Spain\\
$^{5}$ Institut de F\'{\i}sica d'Altes Energies (IFAE), The Barcelona Institute of Science and Technology, Campus UAB, 08193 Bellaterra (Barcelona) Spain\\
$^{6}$ Universit\"ats-Sternwarte, Fakult\"at f\"ur Physik, Ludwig-Maximilians Universit\"at M\"unchen, Scheinerstr. 1, 81679 M\"unchen, Germany\\
$^{7}$ Cerro Tololo Inter-American Observatory, National Optical Astronomy Observatory, Casilla 603, La Serena, Chile\\
$^{8}$ Department of Physics \& Astronomy, University College London, Gower Street, London, WC1E 6BT, UK\\
$^{9}$ Department of Physics and Electronics, Rhodes University, PO Box 94, Grahamstown, 6140, South Africa\\
$^{10}$ Fermi National Accelerator Laboratory, P. O. Box 500, Batavia, IL 60510, USA\\
$^{11}$ CNRS, UMR 7095, Institut d'Astrophysique de Paris, F-75014, Paris, France\\
$^{12}$ Sorbonne Universit\'es, UPMC Univ Paris 06, UMR 7095, Institut d'Astrophysique de Paris, F-75014, Paris, France\\
$^{13}$ Department of Physics and Astronomy, University of Pennsylvania, Philadelphia, PA 19104, USA\\
$^{14}$ Kavli Institute for Particle Astrophysics \& Cosmology, P. O. Box 2450, Stanford University, Stanford, CA 94305, USA\\
$^{15}$ SLAC National Accelerator Laboratory, Menlo Park, CA 94025, USA\\
$^{16}$ Laborat\'orio Interinstitucional de e-Astronomia - LIneA, Rua Gal. Jos\'e Cristino 77, Rio de Janeiro, RJ - 20921-400, Brazil\\
$^{17}$ Observat\'orio Nacional, Rua Gal. Jos\'e Cristino 77, Rio de Janeiro, RJ - 20921-400, Brazil\\
$^{18}$ Department of Astronomy, University of Illinois, 1002 W. Green Street, Urbana, IL 61801, USA\\
$^{19}$ National Center for Supercomputing Applications, 1205 West Clark St., Urbana, IL 61801, USA\\
$^{20}$ Institute of Cosmology \& Gravitation, University of Portsmouth, Portsmouth, PO1 3FX, UK\\
$^{21}$ School of Physics and Astronomy, University of Southampton,  Southampton, SO17 1BJ, UK\\
$^{22}$ George P. and Cynthia Woods Mitchell Institute for Fundamental Physics and Astronomy, and Department of Physics and Astronomy, Texas A\&M University, College Station, TX 77843,  USA\\
$^{23}$ Department of Physics, IIT Hyderabad, Kandi, Telangana 502285, India\\
$^{24}$ Department of Astronomy, University of Michigan, Ann Arbor, MI 48109, USA\\
$^{25}$ Department of Physics, University of Michigan, Ann Arbor, MI 48109, USA\\
$^{26}$ Kavli Institute for Cosmological Physics, University of Chicago, Chicago, IL 60637, USA\\
$^{27}$ Instituto de F\'isica Te\'orica IFT-UAM/CSIC Universidad Aut\'onoma de Madrid, Cantoblanco 28049, Madrid, Spain\\
$^{28}$ Institute of Astronomy, University of Cambridge, Madingley Road, Cambridge CB3 0HA, UK\\
$^{29}$ Kavli Institute for Cosmology, University of Cambridge, Madingley Road, Cambridge CB3 0HA, UK\\
$^{30}$ Astronomy Department, University of Washington, Box 351580, Seattle, WA 98195, USA\\
$^{31}$ Australian Astronomical Observatory, North Ryde, NSW 2113, Australia\\
$^{32}$ Departamento de F\'{\i}sica Matem\'atica,  Instituto de F\'{\i}sica, Universidade de S\~ao Paulo,  CP 66318, CEP 05314-970, S\~ao Paulo, SP,  Brazil\\
$^{33}$ Jodrell Bank Center for Astrophysics, School of Physics and Astronomy, University of Manchester, Oxford Road, Manchester, M13 9PL, UK\\
$^{34}$ Department of Astrophysical Sciences, Princeton University, Peyton Hall, Princeton, NJ 08544, USA\\
$^{35}$ Instituci\'o Catalana de Recerca i Estudis Avan\c{c}ats, E-08010 Barcelona, Spain\\
$^{36}$ Excellence Cluster Universe, Boltzmannstr.\ 2, 85748 Garching, Germany\\
$^{37}$ Faculty of Physics, Ludwig-Maximilians-Universit\"at, Scheinerstr. 1, 81679 Munich, Germany\\
$^{38}$ Max Planck Institute for Extraterrestrial Physics, Giessenbachstrasse, 85748 Garching, Germany\\
$^{39}$ Department of Physics and Astronomy, Pevensey Building, University of Sussex, Brighton, BN1 9QH, UK\\
$^{40}$ Universidade Federal do ABC, Centro de Ci\^encias Naturais e Humanas, Av. dos Estados, 5001, Santo Andr\'e, SP, Brazil, 09210-580\\

\bsp	
\label{lastpage}
\end{document}